\documentclass[journal, draftclsnofoot,onecolumn,12pt]{IEEEtran}

\usepackage[T1]{fontenc}% optional T1 font encoding

\usepackage{cite}
\usepackage[colorlinks,
           linkcolor=red,
           anchorcolor=blue,
           citecolor=green
           ]{hyperref}
\ifCLASSINFOpdf
\else
\fi
\usepackage{amsmath}
\usepackage{amssymb }
\usepackage{amsfonts}

\usepackage{graphicx}
\usepackage{lettrine}
\usepackage{epstopdf}
\usepackage{textcomp,booktabs}
\usepackage[usenames,dvipsnames]{color}
\usepackage{colortbl}
\definecolor{mygray}{gray}{.9}
\definecolor{mypink}{rgb}{.99,.91,.95}
\definecolor{mycyan}{cmyk}{.3,0,0,0}
\usepackage{pifont}
\usepackage{subfigure}
\usepackage{multirow}
\usepackage{url}
\usepackage{color}
\usepackage{bm}

\usepackage{algorithm} %format of the algorithm
\usepackage{algorithmic} %format of the algorithm
\usepackage{multirow} %multirow for format of table

\usepackage[dvipsnames,usenames]{color}
\usepackage[usenames,dvipsnames]{color}

\definecolor{light-gray}{gray}{0.90}
\begin{document}
\setlength{\textfloatsep}{10pt}
\title{Meta Learning-based MIMO Detectors: Design, Simulation, and Experimental Test}
\author{\IEEEauthorblockN{Jing Zhang, Yunfeng He, Yu-Wen Li, Chao-Kai Wen, and Shi Jin}
   \thanks{J. Zhang, Y. He, and S. Jin are with the National
    Mobile Communications Research Laboratory, Southeast University, Nanjing 210096, China (e-mail: jingzhang@seu.edu.cn; heyunfeng@seu.edu.cn; jinshi@seu.edu.cn).}
   \thanks{Y.-W. Li and C.-K. Wen are with the Institute of Communications Engineering, National Sun Yat-sen University, Kaohsiung 80424, Taiwan (e-mail: win912153@gmail.com; chaokai.wen@mail.nsysu.edu.tw).}}

% make the title area\emph{}
\maketitle

\vspace{-0.75cm}
\begin{abstract}
Deep neural networks (NNs) have exhibited considerable potential for  efficiently balancing the performance and complexity of multiple-input and multiple-output (MIMO) detectors.
However, existing NN-based MIMO detectors are difficult to be deployed in practical systems because of their slow convergence speed and low robustness in  new environments.
To address these issues  systematically, we propose a receiver framework that enables efficient online training by leveraging the following  simple observation:
although  NN parameters should adapt to channels, not all of them are channel-sensitive. In particular, we use a deep unfolded NN structure that represents iterative algorithms in signal detection and channel decoding modules as multi layer deep feed forward networks.
An expectation propagation (EP) module, called EPNet, is established for signal detection by unfolding the EP algorithm and rendering the damping factors trainable.
An unfolded turbo decoding module, called TurboNet, is used for channel decoding. This component decodes the turbo code, where trainable NN units are integrated into
the traditional { max-log-maximum \emph{a posteriori}}  decoding procedure.
We demonstrate  that TurboNet is robust for channels and requires only one off-line training.
Therefore, only a few damping factors in EPNet must be re-optimized online.
An online training mechanism based on meta learning is then developed. Here, the optimizer, which is implemented by long short-term memory NNs, is trained to  update
 damping factors efficiently by using a small training set such that they can quickly adapt to new environments.
Simulation results indicate  that the proposed receiver significantly outperforms traditional receivers and that the online learning mechanism can quickly adapt to new environments.
Furthermore, an over-the-air platform is presented to demonstrate the significant robustness of the proposed receiver in practical deployment.

\end{abstract}
%  keywords
\setlength{\baselineskip}{22pt}
\section{Introduction}

The rapid development of modern information society and mobile devices  has exponentially increased the demand for wireless transmission rates. Multiple-input and multiple-output (MIMO) technology can intensively improve spectral efficiency, and it has been widely applied to modern wireless communication systems \cite{cho2010mimo}. The channel capacity and data throughput of a wireless system can be theoretically increased  when the number of isolated antennas in transmitters and mobile terminals is increased. However, embedding a large number of isolated antennas into smartphones is challenging because of the latter's extremely limited space \cite{MTantenna1,MTantenna2}.
In addition, a compromise between computational complexity and signal detection performance must be made when incorporating MIMO technologies into smartphones \cite{MT2019}.
Among currently available MIMO detectors   \cite{mimoreview2015},  maximum likelihood (ML) detection can best achieve optimal performance. However, the  complexity of ML detection exponentially increases with the number of decision variables. Some suboptimal linear detectors, such as zero-forcing and linear minimum mean square error (MMSE) detectors, have been applied to reduce the  computational complexity of MIMO technology. However,  greater performance degradation compared with that under ML detection has been observed.

%\emph{sss}

Iterative MIMO detectors based on approximate message passing (AMP) \cite{2009AMPorigin} and expectation propagation (EP) \cite{2001EPorigin} algorithms have elicited considerable  attention
because of their excellent performance and moderate complexity \cite{2014MIMOamp,EPSeeger2005,2014EP2,2018EPJ}. AMP-based detectors \cite{2014MIMOamp} can achieve Bayes-optimal performance in large-scale MIMO systems when  the elements of the channel matrix have zero-mean independent and identical  sub-Gaussian distributions. However, AMP-based detectors are fragile in terms of the channel matrix and perform poorly outside the independent and identical sub-Gaussian distributions. Researchers have attempted to solve this problem by deriving EP-based detectors \cite{EPSeeger2005,2014EP2,2018EPJ} through the approximation of  posterior distribution with factorized Gaussian distributions. The resultant detectors generally exhibit  better performance than AMP-based detectors over a broad class of MIMO channel matrices. Detection performance can be further improved by adopting a turbo-detection scheme through iterative information exchange between the signal detector and the channel decoder \cite{2017Santos}. Several turbo receivers that use EP-based detectors have been applied \cite{2017Santos,2018Santos,2019epturbo}. However, the convergence speed and stability of such iterative detectors rely heavily on the setting of a number of handcrafted tuning factors with low efficiency.

An emerging technology, called model-driven neural networks (NNs) \cite{2018modeldriven} or deep unfolded NNs \cite{2019unfolding}, exhibits  promising potential for designing MIMO detectors. This technology is used by combining iterative algorithms with tools from NNs to  balance the performance and complexity of MIMO detectors efficiently. The authors of \cite{2019detnet} proposed DetNet to realize MIMO detection by unfolding the iterations of a projected gradient descent algorithm into an NN. In \cite{2019parallelDetnet},   a new DetNet-based detection network, called parallel detection network, was presented to improve the performance of DetNet by dividing a single DetNet into multiple parts and then arranging them in parallel.
In \cite{2019mmnet},  MMNet was introduced by developing an unfolding NN based on the AMP algorithm and by adding a large number of trainable parameters to track different channel realizations.
However, these NNs typically  have an excessive number of training parameters. Model-driven NNs that train fewer parameters than those described in \cite{2019detnet,2019parallelDetnet,2019mmnet} have  been recently developed. For example, a trainable iterative soft thresholding algorithm  \cite{TISTA2019} unfolds an orthogonal AMP (OAMP) algorithm \cite{2017OAMP} and trains only a few parameters through deep learning to solve the problem of sparse signal recovery.
Motivated  by \cite{TISTA2019},  the authors of \cite{DBLP:journals/corr/abs-1809-09336} developed OAMPNet to solve signal detection in MIMO systems \cite{DBLP:journals/corr/abs-1809-09336} and cyclic prefix (CP)-free orthogonal frequency-division multiplexing (OFDM) \cite{2019conf} channels by introducing additional trainable parameters. In \cite{2019EPNet}, EP was unfolded  in an inner detection loop to obtain a deep detection network with learnable damping factors in a turbo receiver.\footnote{We note \cite{2019EPNet} at the final writing stage of this paper. Although our proposed turbo receiver is similar to that of \cite{2019EPNet}, our work differs in terms of the following three points: First, \cite{2019EPNet} does not describe the type of online learning representing the main focus of this paper. Second, no over-the-air platform is presented in \cite{2019EPNet} to verify the robustness of the proposed turbo receiver. Third, we provide interpretable characteristics for the learned parameters in detail.}

Despite these advancements, however, current NN-based MIMO detectors \cite{TISTA2019,DBLP:journals/corr/abs-1809-09336,2019conf,2019EPNet} are difficult to be used in practical environments because the learned parameters are highly  related to specific channel realizations and signal-to-noise ratios (SNRs). NNs must be re-trained when channel realizations or SNRs change. An off-line training mechanism that discretizes all channel realizations and SNRs may be impractical if not impossible. Therefore, an efficient online training mechanism that can quickly adapt to channel realizations and SNRs is crucial for NN-based MIMO detectors. Several  online training mechanisms for wireless communications \cite{2019mmnet,2018jpw,2019viterbinet} have been studied over the last few years. In MMnet \cite{2019mmnet}, a trained model for one channel realization can function  as a strong initialization for training adjacent channel realizations to accelerate online training in OFDM-MIMO systems because of the temporal and spectral correlations in practical MIMO channels. However, MMNet must repeat the complete  algorithm training process at all time intervals because its NN parameters depend on each channel realization. In addition,  training convergence speed remains extremely slow for the first channel realization at each time interval. SwitchNet \cite{2018jpw} was proposed to train several sets of NN parameters off-line on the basis of representative indoor or outdoor channels and to learn how to automatically switch to different sets through online environment sensing. However, SwitchNet is ineffective when the channel environments  completely differ from the training sets. ViterbiNet \cite{2019viterbinet} integrates NNs into the Viterbi algorithm to detect symbols without instantaneous channel state information. An online mechanism is  presented to collect reliable training data as labels when the bit error ratio (BER) is lower than a setting threshold for tracking time-varying channels. However, ViterbiNet may experience severe latency when collecting sufficient  training data online, particularly  in low-SNR regimes. Overall, these online training mechanisms remain unsuitable for practical MIMO systems.

Two important issues must be solved to effectively realize an online training mechanism. First,  training labels must be easily acquired. Second, the convergence speed for training must be sufficiently fast. We consider a specific model-driven NN-based MIMO detector that can be trained  on the basis of  channel statistics rather than channel realizations to solve the first problem. A meta learning technique is introduced to solve the second issue. Meta learning studies the increase in efficiency of learning systems through experience. The goal is to understand how learning can become flexible in accordance with the investigated tasks \cite{2001metasurvey}. A meta learning system should include a learning subsystem that adapts with experience. Experience is gained by utilizing  the meta-knowledge extracted in a previous learning episode on a single dataset and from different domains or problems. Meta learning has been applied to the  field of communications \cite{2019meta,2019demodulate}.

{ The contributions of this current study  are summarized as follows.
\begin{itemize}
\item An unfolded turbo receiver that includes several signal detection and channel decoding modules with an alternating arrangement is proposed in this work. Signal detection is realized in the receiver by representing the EP algorithm as multi-layer deep feed forward networks to optimize the necessary damping factors. The proposed NN-based MIMO detector is called EPNet. Appropriate  damping factors enable the EP algorithm to achieve good detection performance by using only a few iterations and reducing detection complexity. The training labels can be obtained by generating similar channels locally at the receiver side because the damping factors are relevant to the channel statistics rather than instantaneous channel realizations.
\item  Channel decoding relies on a model-driven NN, called TurboNet \cite{2019turbonethe}, to decode the turbo code. Although TurboNet is trained in an additive white Gaussian noise (AWGN) channel and low SNR, it is robust to Rayleigh fading and practical channels. Therefore, TurboNet is only required  to be trained once via an off-line setting, and only EPNet should  be trained online.
\item  A meta learning strategy is used to train the damping factors of EPNet to improve  convergence speed and quickly adapt to new environments. In particular, a long short-term memory (LSTM) optimizer is first trained off-line to learn the rules of gradient descent more efficiently compared with traditional optimizers, such as an adaptive moment estimator (i.e., Adam) \cite{2014adam}. Then, the LSTM optimizer is used to update the damping factors in EPNet online.
\item The combination of EPNet and TurboNet can achieve excellent detection performance and be easily deployed online in practical MIMO systems. In addition to simulations, an over-the-air (OTA) platform is built to verify the effectiveness of the online training mechanism.
\end{itemize}

}

{\bf Notations}---Column vectors are denoted by boldface letters. Superscripts ${{(\cdot )}^{T}}$ and ${{(\cdot )}^{H}}$ represent the transpose and conjugate transpose, respectively.  $\mathbf{I}$ denotes the identity matrix.  The Euclidean norm is denoted by $\left\| \cdot  \right\|$. The expectation operator is represented as $\mathbb{E}\{\cdot \}$. $\mathbb{V}\{\cdot \}$ denotes the covariance matrix. $\mathcal{N}(z;0,{{\sigma }^{2}})$ indicates real-valued Gaussian random variable $z$ with  zero mean and variance ${{\sigma }^{2}}$. Similar to $\mathcal{N}(z;0,{{\sigma }^{2}})$, $\mathcal{N}_{\mathbb{C}}(z;0,{{\sigma }^{2}})$ represents the complex-valued one. The real and imaginary parts of a complex number are represented by ${\rm Re}\{\cdot \}$ and ${\rm Im}\{\cdot \}$, respectively.
\setlength{\baselineskip}{20pt}
\section{System Model}

\begin{figure} [b]
\setlength{\abovecaptionskip}{-1cm}
\setlength{\belowcaptionskip}{0cm}
  \centering
  \includegraphics[width=6.8in]{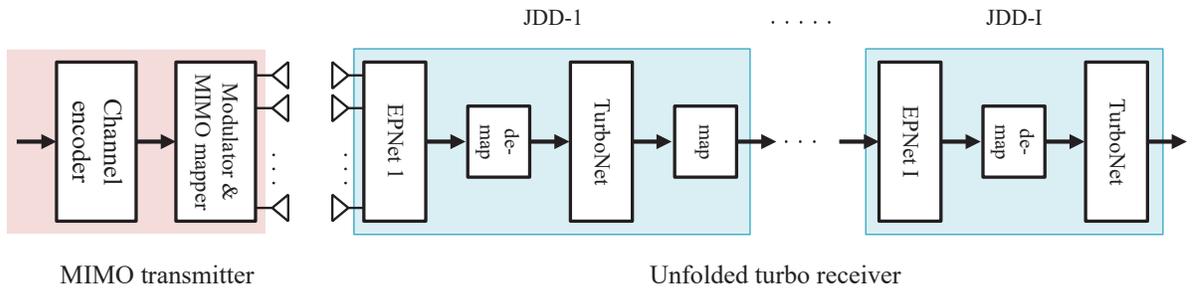}
        \caption{Block diagram of the MIMO system, including the transmitter, channel, and unfolded turbo receiver. The detector and the channel decoder integrate  NNs into traditional algorithms. The unfolded turbo receiver alternately exchanges information between EPNet and TurboNet.}
  \label{system}
 \end{figure}

We consider a MIMO system with $N_t$ transmit and $N_r$ receive antennas. This system has a block digram that  includes the MIMO transmitter and the unfolded turbo receiver, as illustrated in Fig. \ref{system}.

 %\vspace{-1em}

\subsection{Transmitter and Channel}

At the transmitter, the binary information sequence, $\mathbf{a}=[ {{a}_{1}},\ldots ,{{a}_{K}} ]_{{}}$ with ${{a}_{k}}\in \{ 0,\text{ }1 \}$, is encoded by the turbo encoder into a coded bit vector $\mathbf{b}=[b_{1}, \dots, b_{V}]$ with a code rate equal to $R\text{ }=\frac{K}{V}$. The channel encoder contains two identical recursive systematic convolutional encoders (RSCEs). The generator matrix of the RSCE is $[1, g_{1}(D) / g_{0}(D)]$, { where $D$ denotes the  delay unit, generator polynomial $g_{1}(D)=1+D+D^{3}$ indicates the forward path, and  generator polynomial $g_{0}(D)=1+D^{2}+D^{3}$  represents  the feedback path \cite{turboencoder}.} Binary information sequence $\mathbf{a}$ is  transmitted directly as systematic bits $\mathbf{b}^{\rm s}$. The first RSCE generates a sequence of parity bits $\mathbf{b}^{1 {\rm p}}$ from the systematic bits, and the second RSCE generates a sequence of parity bits $\mathbf{b}^{2 {\rm p}}$ from an interleaved sequence of the systematic bits. Then, the codeword, $\mathbf{b}={{[ \mathbf{b}^{\rm s},\mathbf{b}^{1 {\rm p}},\mathbf{b}^{2 {\rm p}} ]}}$, is partitioned into $N$ blocks with length $Q=\log_{2}( M )$, $\mathbf{b}={{[ {{\mathbf{b}}_{1}},\ldots ,{{\mathbf{b}}_{{N}}} ]}}$ with $\mathbf{b}_{n}=[b_{n, 1}, \dots, b_{n, Q}]$, and modulated with a complex $M$-ary quadrature amplitude modulation (QAM) constellation $\mathcal{A}$ with  size $|\mathcal{A}|=M$. Here, these modulated symbols, $\mathbf{x}={{[ {{x}_{1}},\ldots ,{{x}_{N}} ]}}$ with ${{x}_{n}}=\operatorname{Re}\{ {{x}_{n}} \}+j\operatorname{Im}\{ {{x}_{n}} \}\in \mathcal{A}$,  are partitioned into $P$ blocks with  length $N_t$, where $\mathbf{x}={{[ {{\mathbf{x}}{[1]}},\ldots ,{{\mathbf{x}}[P]} ]}}$ with ${{\mathbf{x}}[p]}={{[ {{x}_{p,1}},\ldots ,{{x}_{p,{{N}_{{t}}}}} ]}}^{T}$. Each block is de-multiplexed into $N_t$ substreams through a  serial-to-parallel converter and transmitted to the MIMO channel.
The transmitted symbol energy and energy per bit are denoted as ${{E}_{s}}$ and ${{E}_{b}}$, respectively. $N_0$ indicates the noise spectral density.

\subsection{Principles of Unfolded MIMO Turbo Receiver}

Joint signal detection and channel decoding (JDD) iteration is referred to as ``turbo receiver'' in this paper. The unfolded MIMO turbo receiver is based on the idea of unfolding the traditional turbo receiver \cite{2017Santos,2018Santos}  using a deep NN (DNN) \cite{TISTA2019}. As shown in Fig. \ref{system}, the unfolded MIMO turbo receiver consists of $I$ unfolded JDD modules (called {\tt JDD}$-1$, {\tt JDD}$-2$, $\ldots$, {\tt JDD}$-I$). This unfolding aims to introduce a feed forward network into the signal detection and channel decoder modules. For signal detection, the EP algorithm (hereafter referred to as EP) is used and unfolded as a feed forward network by introducing learnable parameters. The network is called EPNet. An existing NN-based module, i.e., TurboNet \cite{2019turbonethe}, is adopted for channel decoding. Here, log-likelihood ratios (LLRs) that indicate the reliability information of coded bits $\mathbf{b}$  are forwarded between the signal detector and the channel decoder.
The received signal for the $p$-th block is denoted as
\begin{equation}\label{received-signal}
    \mathbf{y}[p]=\mathbf{H}[p] \mathbf{x}[p]+\mathbf{n}[p],
\end{equation}
where $\mathbf{H}[p] \in {\mathbb C}^{N_r \times N_t} $ and $\mathbf{n}[p] \in {\mathbb C}^{N_r} $ represent the MIMO channel matrix and the noise vector, respectively.
Notably, the received noise variance for each received antenna should be different in a practical environment but can be normalized. Therefore, the noise vector follows $\mathbf{n}[p]  \sim \mathcal{N}_{\mathbb{C}}(0, \mathbf{I}_{N_r})$ after whiting, and $\mathbf{H}[p]$ denotes the \emph{equivalent} MIMO channel matrix accordingly. In the remainder of this paper, $p$ is omitted for  ease of notation.
In this study, we assume that the channel matrix $\mathbf H$ is  known at the receiver.

\emph{A posterior} probability that uses the  channel model \eqref{received-signal} is expressed as
\begin{equation}\label{posterior}
    p(\mathbf{x}|\mathbf{y})=\frac{p(\mathbf{y}|\mathbf{x})p(\mathbf{x})}{p(\mathbf{y})}\propto \mathcal{N}_{\mathbb{C}}(\mathbf{y};\mathbf{Hx},\mathbf{I}_{N_r}) {\left(\prod\limits_{n=1}^{N_t}{{{p}_{\sf a}}({{x}_{n}})}\right)},
\end{equation}
where ${p}_{\sf a}({{x}_{n}})$ represents the \emph{a priori} probability density function (pdf) of $\mathbf{x}$. In {\tt JDD}$-1$, ${p}_{\sf a}({{x}_{n}})$ is initialized as the uniform distribution, which can be expressed as
{
\begin{align}
{p}_{\sf a}({{x}_{n}})=\frac{1}{M} \sum_{x \in \mathcal{A}} \delta\left(x_{n}-x\right).
\end{align}
}
In the remaining JDD blocks, ${p}_{\sf a}({{x}_{n}})$ can be constructed through the channel decoding module.
 { The MMSE criterion can be achieved}   by computing
\emph{a posteriori} pdf $p(\mathbf{x}|\mathbf{y})$. However, the direct
computation of $p(\mathbf{x}|\mathbf{y})$ is intractable because of the calculation of the high-dimensional
integral. This situation has motivated us to pursue EPNet to  approximate \eqref{posterior} effectively, as discussed  in detail  in Section III.

In each JDD module, EPNet computes extrinsic pdf $p_{\sf e}(\mathbf{x}| \mathbf{y})$, which is then demapped as  extrinsic LLRs \cite{2017Santos}
\begin{equation}\label{demappedLLR}
    L_{\sf e}(b_{n, q})=\log \frac{\sum_{x_{n} \in \mathcal{A} | b_{n, q}=0} p_{\sf e}(x_{n} | \mathbf{y})}{\sum_{x_{n} \in \mathcal{A} | b_{n, q}=1} p_{\sf e}(x_{n} | \mathbf{y})},
\end{equation}
where $\sum_{x_{n} \in \mathcal{A} | b_{n, q}=0}$ represents the constellation set when the corresponding $q$-th bit of $x_{n}$ is equal to zero. Subsquently, the extrinsic LLRs are delivered to TurboNet for channel decoding. TurboNet combines a  { max-log-maximum \emph{a posteriori} (max-log-MAP)} algorithm with the NNs \cite{2019turbonethe} and provides a new \emph{a priori} LLR $L_{\sf a}(\mathbf{b})$ for EPNet,  the  computation of which  can be found in \cite[(14)]{2019turbonethe}. Then, the new \emph{a priori} LLR is mapped as a new \emph{a priori} pdf, which can be expressed as
\begin{equation}\label{1}
    p_{\sf a}(x_{n} | L_{\sf a}(\mathbf{b}))=\sum_{x \in \mathcal{A}} \delta(x_{n}-x) \prod_{q=1}^{Q} p_{\sf a}(b_{n, q}=\varphi_{q}(x) | L_{\sf a}(\mathbf{b})),
\end{equation}
where $\varphi_{q}(x)$ denotes the $q$-th bit associated with the demapping of symbol $x$. ${p}_{\sf a}({\mathbf{x}})$ is used
by the following the JDD module. This process is repeated for a given maximum number of unfolded layers $I$.

\section{Unfolded Turbo Receiver for MIMO Systems}

An existing TurboNet that is robust for channels with only one \emph{off-line} training is briefly introduced in this section. Then, EPNet is elaborated by unfolding EP and training the damping factors. The damping factors should  be re-optimized when the channel environment changes. Therefore, an online training mechanism is presented to  adjust the damping factors automatically in different environments.

\subsection{TurboNet}

TurboNet \cite{2019turbonethe} is a model-driven NN architecture for turbo decoding that combines an NN with the classic max-log-MAP algorithm \cite{1994maxlogmap}. An original iterative structure \cite{1993turbodecoding} for turbo decoding is unfolded, and each iteration is replaced with  a DNN decoding subnet to design TurboNet. TurboNet integrates a DNN to the max-log-MAP algorithm, or the subnet is obtained by adding trainable parameters into the max-log-MAP algorithm instead of  replacing the entire soft-input soft-output decoder with a black box of fully connected DNN architecture. In particular, a loss function \cite[(16)]{2019turbonethe} is used to evaluate network loss by calculating the mean square  error  between the output of TurboNet and the precise results calculated using the log-MAP algorithm \cite{1997logmap} with given iterations. The trainable parameters of TurboNet can be efficiently learned from the training data, and TurboNet learns to  use systematic and parity information appropriately to provide high error correction capability. In addition, simplified network pruning is performed to preserve several essential training parameters \cite[(15)]{2019turbonethe},  significantly reducing the number of training parameters and improving the robustness of TurboNet and the  error correction capability. The  TurboNet mentioned in the succeeding sections refers to the pruned TurboNet.
%\vspace{-1em}

\subsection{EPNet}

EP \cite{2014EP2,2018EPJ} is an important technique in Bayesian machine learning for approximating posterior beliefs with exponential family distributions. The combination of EP and feed forward NN is applied to signal detection in this subsection.

The complex-valued MIMO system is reformulated into a real-valued system  before EP detection. In the following  context, the equivalent real-valued model is applied. The \emph{a posteriori}  probability of (\ref{posterior}) in a real domain can be rewritten as follows:
\begin{align}\label{pos}
p(\mathbf{{x}}|\mathbf{{y}})=\frac{p(\mathbf{{y}}|\mathbf{{x}})p(\mathbf{{x}})}{p(\mathbf{{y}})} \propto \mathcal{N}({\mathbf{y}};\mathbf{{H}{x}},\mathbf{I}_{2N_r}) \cdot \prod\limits_{n=1}^{2N_t}{{{p}_{\sf a}}({{x}_{n}})} ,
\end{align}
where ${x}_{n}$ for $n=1,2,\ldots, 2N_t$ represents the real/imaginary parts of the modulated symbols. $p(\mathbf{{x}}|\mathbf{{y}})$ is a multi dimensional discrete distribution that maps over a fully connected factor graph because of the likelihood term in (\ref{pos}). The exact inference over $p(\mathbf{{x}}|\mathbf{{y}})$ that is required to evaluate symbol marginals $p(\mathbf{{x}}_{n}|\mathbf{{y}})$ has cost $\mathcal{O}{(M ^{N_{t}} )}$ and becomes infeasible.

EP provides a general-purpose framework for constructing a tractable approximation of $p(\mathbf{{x}}|\mathbf{{y}})$ by using distribution $q(\mathbf{{x}})$ with exponential family distributions\footnote{A comprehensive introduction to exponential families and their properties can be found in\cite{ExponentialFamilies2008}.}. In (\ref{pos}), $\mathcal{N}({\mathbf{y}};\mathbf{{H}{x}},\mathbf{I}_{2N_r})$ belongs to the multivariate Gaussian family ($\mathcal F$) with sufficient statistics $\bm \phi(\mathbf x) = {\phi_1(\mathbf x),\phi_2(\mathbf x),...,\phi_S(\mathbf x)}$, where $\phi_s(\mathbf x)=\{x_i,x_ix_j\}_{i,j=1}^{2N_t}$ \cite{2014EP2}.
The similarity between $p(\mathbf{{x}}|\mathbf{{y}})$  and $q(\mathbf{{x}})$ is achieved by designing $q(\mathbf{{x}})$ to satisfy the moment matching
condition.
\begin{equation}\label{momentmatching}
\mathbb{E}_{q(\mathbf x)} \{ \phi_j(\mathbf x) \} = \mathbb{E}_{p(\mathbf{{x}}|\mathbf{{y}})} \{ \phi_j(\mathbf x) \}, ~j = 1, \ldots , S,
\end{equation}
where $\mathbb{E}_{q(\mathbf x)}$ denotes the expectation with respect to distribution $q(\mathbf x)$.

In the succeeding paragraphs, we present the formulation of the EP updating rules  for polynomial complexity  in accordance with \cite{EPSeeger2005,2018EPJ}. Each  non-Gaussian factor (${{{p}_{\sf a}}({{x}_{n}})}$) in (\ref{pos}) is replaced with an unnormalized Gaussian:
\begin{align}\label{qx}
q(\mathbf{{x}}|\bm{\gamma},\mathbf{\Lambda })\propto \mathcal{N}({\mathbf{y}};\mathbf{{H}{x}},\mathbf{I}_{2N_r}) \cdot \prod\limits_{n=1}^{2N_t}{{{e}^{{{\gamma }_{n}}{{{{x}}}_{n}}-\frac{1}{2}{{\Lambda }_{n}}{x}_{n}^{2}}}}=\mathcal{N}({\mathbf{y}};\mathbf{{H}{x}},\mathbf{I}_{2N_r}) \cdot \mathcal{N}({\mathbf{x}};{{\mathbf{\Lambda }^{-1}}} \bm{\gamma},{{\mathbf{\Lambda }^{-1}}}),
\end{align}
where ${{\gamma }_{n}}\in {\mathbb R}$, ${{\Lambda }_{n}}\in {{\mathbb R}^{+}}$, $\bm{\gamma}=[{\gamma }_{1},...,{\gamma }_{2N_t}]^T$, and $\bm{\Lambda }=\text{diag} ([{\Lambda }_{1},...,{\Lambda }_{2N_t}])$. Initiation is set as ${{\bm{\gamma }}^{(0)}}=\mathbf{0}$ and ${{\mathbf{\Lambda }}^{(0)}}=\frac{1}{2{E}_{s}}\mathbf{I}_{2N_t}$ (this process leads to the MMSE solution \cite{mmse2004}). The mean vector $\bm{\mu }$ and the covariance matrix $\bm{\Sigma}$ of
$q(\mathbf{{x}}|\bm{\gamma},\mathbf{\Lambda })$ in the $l$-th iteration can be derived by applying the Gaussian production lemma\footnote{
The product of two Gaussians gives another Gaussian  \cite{15gaussian},
%$ \mathcal{N}(x;c,C) \cdot \mathcal{N}(x;d,D)
%=\mathcal{N}(0;c-d,C+D) \cdot \mathcal{N}\left(x;\frac{c/C+d/D}{1/C+1/D},\frac{1}{1/C+1/D}\right)$.
\begin{equation*}
 \mathcal{N}(x;c,C) \cdot \mathcal{N}(x;d,D)
 =\mathcal{N}(0;c-d,C+D) \cdot \mathcal{N}\left(x;\frac{c/C+d/D}{1/C+1/D},\frac{1}{1/C+1/D}\right).
\end{equation*}
} to (\ref{qx})   as follows:
\begin{equation}\label{average}
 {{\bm{\Sigma}}^{(l)}}={{\left( {{{\mathbf{{H}}}}^{T}}\mathbf{{H}}+  \mathbf{\Lambda }^{(l-1)} \right)}^{-1}}, ~~~~
 {{\bm{\mu }}^{(l)}}={{\bm{\Sigma}}^{(l)}}\left( {{{\mathbf{{H}}}}^{T}}\mathbf{{y}}+\bm{\gamma}^{(l-1)} \right),
  \end{equation}
respectively. The EP iterative method approximates the solution for (\ref{momentmatching}) by recursively updating pairs $(\bm{\gamma},\bm{\Lambda})$ in (\ref{average}). $l$ denotes the $l$-th iteration of EP, and the maximum number of iterations is set to $L$.
\begin{figure}[t]
\setlength{\abovecaptionskip}{-0.2cm}
\setlength{\belowcaptionskip}{-0.5cm}
  \centering
  \includegraphics[width=4.5in]{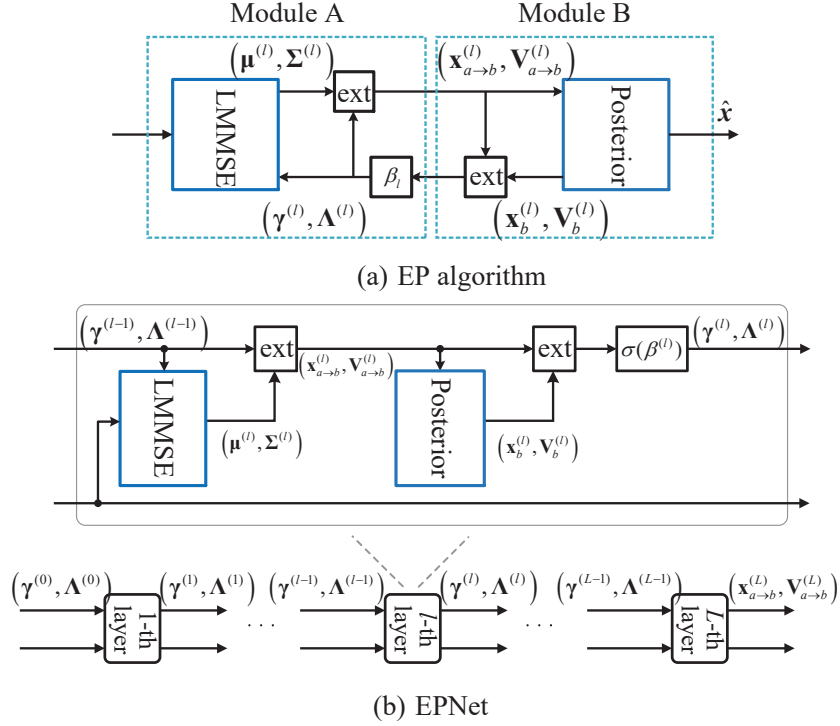}
        \caption{EP is unfolded as EPNet.}
  \label{EPnet}
\end{figure}

The diagram of EP is provided  in Fig. \ref{EPnet}(a), which shows the iteration between modules A and B. The  passing of message from A to B is denoted as ${a\to b}$, and  the reverse is denoted as ${b\to a}$.  The pair $(\bm{\gamma},\bm{\Lambda})$ can be computed as follows:
\begin{itemize}

\item[1)] The cavity marginal $q^{(l) \textbackslash n}(x_n)={q^{(l)}(x_n|\bm{\gamma},\bm{\Lambda})}/{{{e}^{{{\gamma }_{n}}{{{{x}}}_{n}}-\frac{1}{2}{{\Lambda }_{n}}{x}_{n}^{2}}}}$ is computed. \\
Given $q(\mathbf{{x}}|\bm{\gamma},\mathbf{\Lambda })\sim \mathcal{N}({\mathbf{x}};{{\bm{\mu }}^{(l)}},{{\bm{\Sigma}}^{(l)}})$,  the extrinsic covariance matrix $\mathbf{{V}}_{a\to b}^{(l)}$ and  the mean matrix $\mathbf{{x}}_{a\to b}^{(l)}$ of the  cavity marginal can be derived as
\begin{equation}
  \left\{
   \begin{aligned}
  &\mathbf{{V}}_{a\to b}^{(l)}={{\left[{\left( {{\bm{\Sigma}}^{(l)}}  \right)^{-1}}- \mathbf{\Lambda }^{(l-1)} \right]}^{-1}},\\
  &\mathbf{{x}}_{a\to b}^{(l)}= \mathbf{{V}}_{a\to b}^{(l)} \left[{\left(  {{\bm{\Sigma}}^{(l)}} \right)^{-1}}{{{\bm{\mu }}^{(l)}}}-\bm{\gamma}^{(l-1)}\right],
   \end{aligned}
   \right.
  \end{equation}

\item[2)] The mean and variance of distribution $\hat p^{(l)}(x_n|\mathbf y) \propto q^{(l) \textbackslash n}(x_n) {p}_{\sf a}({{x}_{n}})$ are computed.\\
The posterior mean matrix $\mathbf{{x}}_{b}^{(l)}$ and the covariance matrix $\mathbf{{V}}_{b}^{(l)}$ of $\hat p^{(l)}(\mathbf {x}|\mathbf y)$ in module B are obtained from $\mathcal{N}({\mathbf{x}};\mathbf{{x}}_{a\to b}^{(l)},\mathbf{{V}}_{a\to b}^{(l)})$ and the real domain constellation, which can be expressed as
 \begin{equation}%\label{eq_nodamping}
  \left\{
   \begin{aligned}
  &\mathbf{{V}}_{b}^{(l)}= \mathbb{V}{\left\{\mathbf{{x}} \big|\mathbf{{x}}_{a\to b}^{(l)},\mathbf{{V}}_{a\to b}^{(l)} \right \}},\\
  &  \mathbf{{x}}_{b}^{(l)}= \mathbb E{\left\{\mathbf{{x}} \big|\mathbf{{x}}_{a\to b}^{(l)},\mathbf{{V}}_{a\to b}^{(l)} \right \}},
   \end{aligned}
   \right.
  \end{equation}
where the expectations are derived from $\hat p^{(l)}(\mathbf x|\mathbf y)$.

\item[3)]   Pair $(\bm{\gamma}^{(l)},\bm{\Lambda}^{(l)})$ is refined with $\mathbb{E}_{q(\mathbf x|\mathbf{\Lambda },\bm{\gamma} )}=\mathbb{E}_{\hat p(\mathbf x|\mathbf y)}$ and $\mathbb{V}_{q(\mathbf x|\mathbf{\Lambda },\bm{\gamma})}=\mathbb{V}_{\hat p(\mathbf x|\mathbf y)}$, and then is updated as
  \begin{equation}\label{eq_nodamping}
  \left\{
   \begin{aligned}
  & \mathbf{\Lambda }^{(l)}=  {\left(\mathbf{{V}}_{b}^{(l)}\right)}^{-1}-{\left(\mathbf{{V}}_{a\to b}^{(l)}\right)}^{-1},  \\
  & \bm{\gamma }^{(l)}=   {\left(\mathbf{{V}}_{b}^{(l)}\right)}^{-1} \mathbf{{x}}_{b}^{(l)} - {\left(\mathbf{{V}}_{a\to b}^{(l)}\right)}^{-1} \mathbf{{x}}_{a\to b}^{(l)} .
   \end{aligned}
   \right.
  \end{equation}
\end{itemize}

Notably, $\bm{\Lambda  }$ in (\ref{qx}) is an inverse variance term and should be positive. However, the parameter updated in (\ref{eq_nodamping}) may return a negative value, resulting in the exploration of the numerical value and convergence failure. In such case, the previous values should be kept for those parameters \cite{2017thesis}. In addition, parameter updating must be smoothed through a convex combination with the former value to improve the robustness of the algorithm \cite{2014EP2}; that is,

\begin{equation}\label{eq_damping}
  \left\{
   \begin{aligned}
&\bm \Lambda ^{(l)}\leftarrow\beta^{(l)} \bm \Lambda ^{(l)}+(1-\beta^{(l)} )\bm \Lambda ^{(l-1)},\\
&\bm \gamma ^{(l)}\leftarrow\beta^{(l)} \bm \gamma ^{(l)}+(1-\beta^{(l)} )\bm \gamma ^{(l-1)}.\\
   \end{aligned}
   \right.
  \end{equation}
Damping factors, $\bm \beta=[\beta^{(1)},...,\beta^{(L)}]$, are introduced to prevent the EP solution from exploding.

Updating $(\bm{\gamma}^{(l)},\bm{\Lambda}^{(l)})$ with a damping factor is crucial for maintaining the stability of the EP solution. However,  convergence speed is affected by  damping factors. We first review the parameter updating methods used in previous studies \cite{2018EPJ,2014EP2,2018Santos} and then explain those used in EPNet. In accordance with \cite{2014EP2}, several parameters, including minimum allowed variance $\epsilon$, damping procedure $\bm \beta$, and number of iterations $L$, must be tuned. $\epsilon$ ensures the non negativity of the variance, and $\bm \beta$ determines the stability and convergence speed of the algorithm. The computational complexity of the algorithm is linearly related to $L$. $\epsilon$  gradually decreases to avoid instability caused by fast updates. In \cite{2018EPJ}, for example,  when fast updates are performed (i.e., $\beta^{(l)}=0.95, \forall l \in \{1,\cdots,L\}$),
$\epsilon$  started with a small value during the first four iterations and then exponentially decreased as $1/2^{\max(l-4,1)}$. However,  the authors of \cite{2018Santos} found that fast updates can induce  instability for large modulations and turbo schemes. For the turbo receiver, the damping factors should be re-tuned in every turbo iteration \cite{2017Santos,2018Santos}. In \cite{2018Santos}, $\beta^{(l)}$ started with a conservative value and exponentially increased with the number of turbo iterations $i$, i.e., $\beta^{(l)}=\min( 0.1 \cdot e^{i/1.5}, \,0.7)$. The growth of $\beta^{(l)}$ reduced the number of EP iterations when the turbo procedure was used. The parameter settings used in different methods are listed in Table \ref{table_parameter}. Damping factors $\bm \beta$ are manually tuned in accordance with various scenarios to achieve  a trade-off between convergence speed and stability. The tuning process of the damping factors  is customized and exhibits  low efficiency.

\begin{table} %[!t]
  \centering
  \footnotesize
        \caption{Parameter Settings of  EP for the Signal Detection and JDD.}
  \begin{tabular}{lllllll}    %lcrrr
   \toprule
      Type & Algorithm  & $\epsilon$    &  $ \beta^{(l)}$       & $L$ & $I$\\
   \midrule
   \multirow{2}{*}{Detection}
   &EPD\cite{2014EP2} &$5e^{-7}$ &$0.2$ &$10$ & $0$\\
            &EC\cite{2018EPJ}&$1/2^{\max(l-4,1)}$ &$0.95$ &$10$ & $0$\\
            \hline
            JDD & nuBEP\cite{2018Santos}&$1e^{-8}$  &$\min(0.1 \cdot e^{i/1.5}, \, 0.7)$ &$3$& $5$ \\
   \bottomrule
  \end{tabular}
  \label{table_parameter}
 \end{table}

In contrast with  the traditional EP, damping factors $\bm \beta$ are set as trainable parameters and trained by an unfolded NN, as shown in Fig. \ref{EPnet}(b).
Each layer in Fig. \ref{EPnet}(b) contains one adjustable variable $\beta^{(l)}$. Therefore, the total number of learnable variables is equal to $L$ when $L$ layers are found.
The convergence, stability, and speed of EP can be significantly improved through large data training by using the learnable damping factors. In contrast with other NN-based signal detectors, such as DetNet \cite{2019detnet}, the number of learnable variables of EPNet is independent of  the number of antennas in MIMO systems and only related to the number of layers $L$. This characteristic is advantageous for large-scale MIMO systems. A sigmoid function, $\sigma (\cdot)$, is used to constrain the range of damping factors, i.e., $\sigma (\beta^{(l)} )\in [0,1]$. Then, (\ref{eq_damping}) is reformulated as
\begin{equation}
  \left\{
   \begin{aligned}
  & \mathbf{\Lambda }^{(l)}\leftarrow \sigma (\beta^{(l)} ) \mathbf{\Lambda }^{(l)} +(1-\sigma (\beta^{(l)}  ) )\mathbf{\Lambda }^{(l-1)}\\
  & \bm{\gamma}^{(l)}\leftarrow \sigma (\beta^{(l)}  ) \bm{\gamma}^{(l)} +\left(1-\sigma (\beta^{(l)}  ) \right)\bm{\gamma}^{(l-1)}
   \end{aligned}.
   \right.
  \end{equation}

In {\tt JDD}$-i$, the approximated extrinsic pdf $q_{\sf e}^i\sim \mathcal{N}({\mathbf{x}}; \mathbf{{x}}_{a\to b}^{(L)}, \mathbf{{V}}_{a\to b}^{(L)})$  is outputted by EPNet and demapped as extrinsic LLRs $L_{\sf e}^i$ in accordance with (\ref{demappedLLR}). Then, the extrinsic LLRs are delivered to TurboNet,  outputting the new \emph{a priori} LLRs $L_{\sf a}^i$. Subsequently, $L_{\sf a}^i$ is mapped as \emph{a priori} pdf $p_{\sf a}^i$ to construct the mean and variance of constellation $(\tilde{\mathbf{x}}_i,\widetilde{\mathbf{V}}_i)$. In the {\tt JDD}$-(i+1)$ phase, the initial pair of $(\bm{\gamma}^{(0)}_{i+1},\bm{\Lambda}^{(0)}_{i+1}
),~ i=1, \ldots,I-1$ for the corresponding EPNet is computed as
\begin{equation}\label{eq2_damping}
 \mathbf{\Lambda }^{(0)}_{i+1}\leftarrow   \widetilde{\mathbf{V}}_i^{-1} , ~~~
 \bm{\gamma}^{(0)}_{i+1}\leftarrow  \widetilde{\mathbf{V}}_i^{-1} \, \tilde{\mathbf{x}}_i .
  \end{equation}
In {\tt JDD}$-1$, the initiation is set as ${{\bm{\gamma }}^{(0)}_1}=\mathbf{0}$, ${{\mathbf{\Lambda }}^{(0)}_1}=\frac{1}{2{E}_{s}}\mathbf{I}_{2N_t}$.

In the {\tt JDD}$-i$ of the unfolded turbo receiver, the initialization pair of $(\bm{\gamma}^{(0)}_{i},\bm{\Lambda}^{(0)}_{i})$ inputted into EPNet is more accurate than those inputted into the EPNet in  {\tt JDD}$-(i-1)$. Therefore, the EPNet in {\tt JDD}$-i$ should  readjust the damping factors. For the unfolded turbo receiver, EPNets are sequentially trained  $I$ times. The total number of training parameters of EPNets for the unfolded turbo receiver is ${I L}$.

The number of training parameters of model-driven NN-based methods \cite{DBLP:journals/corr/abs-1809-09336,TISTA2019,2019conf} significantly decreases compared with data-driven NN-based methods \cite{8052521liye,2019detnet}. In particular, EPNet can work in every channel realization with a baseline performance without training by using appropriate  initial values of the damping factors, such as nonuniform block EP (nuBEP) in Table \ref{table_parameter}. Training the damping factors can incrementally improve the baseline performance. However, these ${I L}$ trainable parameters of EPNets must  be re-trained when the channel environment considerably  changes. Consequently, an online training mechanism should be developed to adapt to  environment changes.

%\vspace{-1em}
\subsection{Online Training Mechanism based on Meta Learning}
In this subsection, an online training mechanism based on meta learning is introduced to  learn the optimal damping factors of EPNets quickly.

\begin{figure}[b]
\setlength{\abovecaptionskip}{-0cm}
\setlength{\belowcaptionskip}{-0cm}
  \centering
  \includegraphics[width=5in]{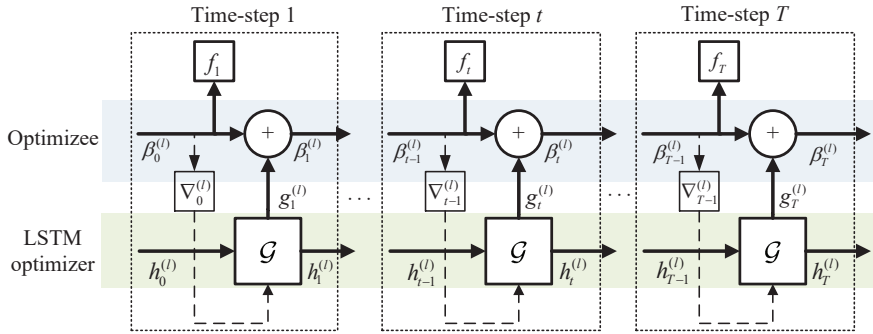}
        \caption{Computational graph between the LSTM optimizer and the single optimizee. }
        %(a) Update of the optimizees using LSTM optimizer for a training epoch; (b) Computational graph used for computing the gradient of the optimizer and updating the optimizee for a single optimizee. }
  \label{LSTMoptimizer}
\end{figure}

In accordance with \cite{2016metalearning}, an optimization algorithm based on meta learning can be modeled as a learning problem, which can be implemented using an LSTM network and replacing traditional optimization algorithms, such as Adam, root mean square prop (RMSprop), and stochastic gradient descent. The new optimizer outperforms these hand-designed optimization algorithms in terms of convergence speed and generalizes well to new tasks. Motivated  by \cite{2016metalearning}, an online training mechanism based on meta learning for EPNet is developed. In particular, we involve the training of two NNs. First, an  LSTM optimizer is developed, as shown in Fig. \ref{LSTMoptimizer}. This optimizer  is implemented using LSTM networks and can realize the function of gradient descent, similar to traditional optimizers.
%All the LSTM networks in Fig. \ref{LSTMoptimizer} have the same structure, and their details are shown in Fig. \ref{LSTMoptimizer}.
Second,  EPNet is trained using the LSTM optimizer, and  the optimizee is set as the damping factor and expected to reach convergence quickly.

%$t$-th
\textbf{Step 1: Learning the LSTM optimizer.}
The LSTM optimizer should be capable of  learning the rules of gradient descent. To achieve this objective, an $L$-dimensional quadratic function is selected  as the objective function of a task, such as the $j$-th task, which is expressed as
\begin{equation}\label{quadratic}
    f(\bm\beta; {\mathbf W}_j, {\mathbf q}_j )={\left\| {\mathbf W}_j{\bm\beta}-{\mathbf q}_j \right\|}^2, ~j=1,...,J,
\end{equation}
where $\mathbf{\mathbf W}_j$ is an $L\times L$ random matrix, and $\mathbf{\mathbf q}_j$  is an $L\times 1$ random vector. Each element of $\mathbf{\mathbf W}_j$  and $\mathbf{\mathbf q}_j$   follows standard normal distributions. $J$ denotes the number of tasks. Quadratic functions with $L$ dimensions are selected because EPNet is involved in $L$  layers. For different ${\mathbf W}_j$ and ${\mathbf q}_j$, the optimal $\bm\beta$ that achieves minimum $ f({\bm\beta}; {\mathbf W}_j, {\mathbf q}_j)$ varies. Therefore, different realizations of $\mathbf{\mathbf W}_j$ and ${\mathbf q}_j$ can constitute different tasks for meta learning.

%Fig. \ref{LSTMoptimizer} shows the update of the optimizees in a training epoch, which amounts to an update of Adam optimizers. The LSTM optimizer is implemented by LSTM networks,  which consists of $T$ LSTM units $\cal{G}$ and updates the corresponding optimizee $T$ time steps. The output in Fig. \ref{LSTMoptimizer} is updated optimizees $\beta_{T}^{l}$ in the epoch.
A computational graph is used to compute the gradient of the optimizee, $\beta^{(l)}$, for the LSTM optimizer, as presented in Fig. \ref{LSTMoptimizer}.  This figure shows an optimizee and an LSTM optimizer with $T$ time steps and $T$ LSTM units, where $t$ denotes the $t$-th time step of the LSTM optimizer; ${\cal G}$ denotes a LSTM unit  with a specific number of hidden layer neurons and layers; $\beta_t^{(l)}$ is the $l$-th updated optimizee parameter in time step $t$; and $f_t$ denotes the value of (\ref{quadratic}) in the $t$-th time step, i.e., $ f_t = f({\bm\beta}_t; {\mathbf W}_j, {\mathbf q}_j )={\|{\mathbf W}_j{\bm\beta}_t-{\mathbf q}_j\|}^2 $. In time step $t$, one input of ${\cal G}$ is the partial derivative, $\nabla_{t}^{(l)}$, which is computed as
\begin{equation}\label{gradient}
\nabla _{t}^{(l)}= \left. \frac{\partial f({\bm \beta} ; {\mathbf W}_j, {\mathbf q}_j) }{ \partial \beta^{(l)}} \right|_{\beta^{(l)} = \beta_{t}^{(l)}}.
\end{equation}
The updating step of the optimizee,  $g_t^{(l)}$, is the output of ${\cal G}$ in which its state is explicitly denoted by $h_{t-1}^{(l)}$. In general, the inputs of ${\cal G}$ in time step $t$ are $\nabla _{t-1}^{(l)}$ and ${h}_{t-1}^{(l)}$, and the outputs are ${g}_{t}^{(l)}$ and ${h}_{t}^{(l)}$. The updating rules for an LSTM unit and the optimizee are expressed as follows:
\begin{equation}\label{updaterule}
  \left\{
   \begin{aligned}
 &\Big( g_{t}^{(l)},h_{t}^{(l)} \Big) ={\cal G} (\nabla_{t-1}^{(l)},h_{t-1}^{(l)}; {\bm \Theta}), \\
&\beta _{t}^{(l)} =\beta _{t-1}^{(l)}+g_{t}^{(l)},
   \end{aligned}
   \right.
  \end{equation}
where ${\bm \Theta}$ represents all the adjustable parameters in LSTM unit ${\cal G}$ and is randomly initialized. The output of the LSTM optimizer is $g_{T}^{(l)}$, as shown in Fig. \ref{LSTMoptimizer}, which is the output of the LSTM unit ${\cal G}$ in the last time step.
In particular, the gradients on  ${\bm \Theta}$ are allowed to flow along solid lines, and  the gradients along  dashed lines are dropped.  The arrows represent the passing of  variables. Disregarding the gradients of ${\bm \Theta}$ along the dashed lines is tantamount  to assuming that the gradient of the optimizee, ${ \nabla }_{t}$,  depends only on $f$ and exhibits  no relationship with ${\bm \Theta}$,  i.e., $\partial {{\nabla }_{t}}/\partial \Theta =0$ \cite{2016metalearning}. Therefore, the LSTM optimizer avoids calculating the second derivative of $f$, i.e., $\frac{{{\partial ^2}f}}{{\partial \beta \partial \Theta }}$, reducing the complexity of training.

All the other optimizees $\bm \beta^{\textbackslash (l)}$ have the same computational graph as $\beta^{(l)}$ and use a structure identical to that in Fig. \ref{LSTMoptimizer}. However, the inputs and the hidden states are separate. The loss function for training the LSTM optimizer is computed as
\begin{equation}\label{loss1}
    \mathcal L({\bm \Theta}) =\frac{1}{JL}\left( \sum\limits_{j=1}^{J}{f(\bm\beta; {\mathbf W}_j, {\mathbf q}_j )} \right).
\end{equation}
The objective of training the LSTM optimizer is to minimize the value of $\mathcal{L}({\bm \Theta} )$ by using the Adam optimizer on ${\bm \Theta}$, which is related to all the optimizees $\bm \beta$.
During an epoch, $\beta^{(l)}$ is { initialized as 1}. Then, $\beta^{(l)}_{1}$ is updated $T$ time by the LSTM optimizer following (\ref{updaterule}), and obtains $\beta _{T}^{(l)} =\beta _{T-1}^{(l)}+g_{T}^{(l)}$. Lastly, $\bm\beta _{T}=[\beta _{T}^{(1)}, \ldots,\beta _{T}^{(L)}]$ is inputted into (\ref{loss1}), and ${\bm \Theta}$ is { updated only once} using the Adam optimizer by minimizing (\ref{loss1}). In the succeeding epochs, we use
$J$ new tasks, {where ${\mathbf W}_j$ and ${\mathbf q}_j$ are regenerated, and $\bm\beta$ is  reinitialized}. The   same process is repeated until the LSTM optimizer achieves convergence. { After the LSTM optimizer reaches convergence, the updating  step of the optimizee, $g_{t}^{(l)}$, is determined by the input of the LSTM unit ${\cal G}$,$\nabla _{t-1}^{(l)}$ and $h_{t-1}^{(l)}$. $g_{t}^{(l)}$ will be positive when $\nabla _{t-1}^{(l)}$ is negative, and $g_{t}^{(l)}$ will be negative when $\nabla _{t-1}^{(l)}$ is positive.}
Although $\nabla _{t}^{(l)}$ in (\ref{gradient}) is only related to the current optimizee $\beta^{(l)}$ and unrelated to the other optimizees $\bm \beta^{\textbackslash (l)}$,  $\mathbf W$ and $\mathbf q$ with $L$ dimensions are adopted. This setting is applied to improve the extensiveness of the trainable parameters ${\bm \Theta}$ of the LSTM optimizer and enables the trained LSTM optimizer to adjust to multiple optimizees in parallel. After training, the parameters of the LSTM unit, ${\bm \Theta}$, are obtained.

The LSTM optimizer can be used to update the optimizees in parallel during the testing phase because it is used to update different optimizees for various  tasks during  the training phase. In addition, the optimizer can be  utilized to train other functions apart from  quadratic functions, such as the loss function in EPNet. During the training of the LSTM optimizer, the dataset originated from the quadratic functions of (\ref{quadratic}). Although task \eqref{quadratic} is nonidentical to that of EPNet, the LSTM optimizer can still provide better gradient descent than the traditional optimization algorithm for EPNet. This funding  is attributed to  the use of recurrence allowing the LSTM optimizer to learn dynamic updating rules that integrate information from the history of gradients, which is similar to momentum.

% The LSTM optimizer is efficient for EPNet. The LSTM optimizer generalizes well to new tasks.

\textbf{Step 2: Training the damping factors of EPNet by using the LSTM optimizer.}
The LSTM optimizer is utilized to train the damping factors of EPNet to  achieve convergence quickly and increase the efficiency of online training. The loss on all the  $L$ layers is defined as the loss function in training EPNet, which is expressed as follows\footnote{The loss function is different from that in \cite{2019EPNet}.}:
\begin{equation}\label{loss-ep}
    f_{\rm EP}(\bm \beta)=\frac{1}{L} \sum_{l=1}^{L}\left\|{\mathbf{x}}_{a\to b}^{(l)}-{\mathbf{x}}\right\|_{2}^{2}.
\end{equation}
The updating rule for $\bm \beta$ is determined by the LSTM optimizer, which has been trained in Step 1. The LSTM optimizer minimizes a new task $f_{\rm EP}(\bm \beta)$ rather than \eqref{quadratic}.\footnote { Note that $f$ in Fig. \ref{LSTMoptimizer} is replaced by $f_{\rm EP}(\bm \beta)$ in this step.}

The online training mechanism begins  to train the damping factors of EPNets when the channel condition considerably  changes, such as the mobile terminal moving from indoor to outdoor. If the online training phase is triggered, then training labels $(\mathbf x,\mathbf y)$ are generated at the receiver side in accordance with  \eqref{received-signal}, where
the modulated symbols $\mathbf{x}$ are randomly generated given the transmitted constellation and the MIMO channel matrices $\mathbf H$ are based on those
acquired during the previous period.\footnote{Notice that the noise vector follows the standard Gaussian distribution since the channel matrices have been normalized.} Subsequently, the damping factors of EPNets are  trained by the LSTM optimizer by using the generated dataset. The damping factors for each EPNet are utilized online,  enabling the unfolded turbo receiver to improve the performance of the traditional turbo receiver.

\section{Simulation Results}

This section presents the simulation  results of the proposed unfolded turbo receiver for MIMO systems. The detailed parameter settings are first described, and then the performance of EPNet and TurboNet is  analyzed. Lastly, the unfolded turbo receiver is evaluated.

\subsection{Parameter Setting}

In the simulation, an ${8 \times 8}$ or ${32 \times 32}$ MIMO system is evaluated. The modulation is set to quadrature phase-shift keying (QPSK), 16-QAM, and 64-QAM. The channel matrix $\mathbf{H}$ is sampled from an independent and identical  Gaussian distribution (i.e., each colume of $\mathbf{H}$ follows the distribution  $ \mathcal{N}_{\mathbb{C}}(0,(1 / N_r) \mathbf{I}_{N_r})$). For signal detection, the initiation parameters follow Table \ref{table_parameter} for JDD. For channel decoding, the turbo code is used, where the message bit length $K$ is set to 40 or 120 and encoded using  the convolutional encoder at a rate of $1/2$.

TurboNet and the LSTM optimizer are first trained \emph{off-line}. Then, EPNets are trained by the LSTM optimizer \emph{online}. TurboNet is trained by minimizing the cost \cite[(16)]{2019turbonethe} with the Adam optimizer at a learning rate of $10^{-5}$. The LSTM unit $\cal G$ is realized by two-layer LSTMs with five hidden units in each layer. time step $T$ is set to 20. The LSTM optimizer is trained by minimizing $\mathcal L$ in \eqref{loss1} with 20 tasks. Minimization is performed using Adam with a learning rate of $10^{-3}$. EPNet is trained by the LSTM optimizer with generated labels. Training data $(\mathbf{x, y})$ are generated by using  (\ref{received-signal}) with $5,000$ channel realizations, and $f_{\rm EP}$ is selected as the cost function. The training parameters of TurboNet, the LSTM optimizer, and EPNet are summarized in Table \ref{train1}.

\begin{table}[!ht]
  \centering
  \caption{Training Parameters of EPNet. }
  \footnotesize
  \begin{tabular}{>{\sf }lll|l|ll}    %
   \toprule
    &Type&TurboNet&LSTM optimizer & EPNet \\
   \midrule
   &Loss function &\cite[(16)]{2019turbonethe}&$\mathcal L$&   $f_{\rm EP}$\\
            &Batch size &   150 &1&   5000\\
            &Tasks size &   N/A &20&   N/A\\
   &Epoch &  500&100&  100\\
   &Learning rate &   $10^{-5}$&$10^{-3}$& N/A  \\
   &Optimizer &   Adam&   Adam&   LSTM optimizer\\
   \bottomrule
  \end{tabular}
  \label{train1}
 \end{table}

The following naming conventions are used to  present the performance concisely, where EP and EPNet are outputs for an uncoded MIMO system and the others are outputs  for turbo receivers (iteratively exchange LLRs between the signal detector and the channel decoder):

\begin{itemize}
\item EP ($L=l$): EP with $l$ iterations, where the damping factor is set as  0.1 for each iteration.

\item EPNet ($L=l$): EPNet with $l$ layers, where the damping factors are trained by the LSTM optimizer.

\item EP+Turbo ($I=i$): EP for signal detection, and the max-log-MAP algorithm for channel decoding with $i$ iterations between the detector and the decoder.

\item EP+TurboNet ($I=i$): EP for signal detection, and TurboNet for channel decoding with $i$ iterations between the detector and decoder.

\item EPNet+Turbo ($I=i$): EPNet for signal detection, and the max-log-MAP algorithm for channel decoding with $i$ iterations between the detector and decoder.

\item EPNet+Turbo-log ($I=i$): EPNet for signal detection, and log-MAP algorithm for channel decoding with $i$ iterations between the  detector and the decoder.

\item EPNet+TurboNet ($I=i$): EPNet for signal detection and TurboNet for channel decoding with $i$ iterations between the detector and the decoder.

\end{itemize}
We use $L=5$ and $I=4$ when no special illustrations are found.

The energy per bit to noise power ratios in the uncoded and coded systems are denoted as
$E_B/N_0$ and $E_b/N_0$, respectively.\footnote{ For the MIMO system with $M$-QAM constellation, the relationship between $E_b/N_0$ and $E_s/N_0$ is expressed as $E_s/N_0= E_b/N_0+10\log\left(\log_2(M) \right)$. The relationship between $E_B/N_0$ and $E_b/N_0$ is expressed as $E_B/N_0= E_b/N_0 +10\log(1/R )$.} In this section, we assume that the received $E_B/N_0$ or $E_b/N_0$ in each antenna is identical to make the simulation results intuitive.

\subsection{Performance Analysis of EPNet and TurboNet}
\begin{table}[!t]
  \centering
  \footnotesize
        \caption{ Performance Comparison and Damping Factors Trained by Meta Learning with 16-QAM in Rayleigh Channel.}
  \begin{tabular}{>{\sf }lll|l|l|l|l}    %lcrrr
   \toprule
    &$E_B/N_0$& -1 dB    &  4 dB       & 9 dB & 14 dB & 19 dB \\
                      %&(W)&  (W)  & (S) & (S) \\
   \midrule
               BER&EP ($L=5$) &$2.8430e^{-1}$ &$1.8092e^{-1}$ &$9.7812e^{-2}$ &$1.4781e^{-2}$  &$6.7563e^{-4}$\\
            &EP ($L=10$) &$2.7954e^{-1}$ &$1.7729e^{-1}$ &$8.3663e^{-2}$ &$8.8561e^{-3}$ &$3.2534e^{-4}$\\
            &EP ($L=15$)&$2.7973e^{-1}$ &$1.7737e^{-1}$ &$8.1863e^{-2}$ &$7.5135e^{-3}$ &$2.3675e^{-4}$\\
            &EPNet ($L=5$) &$\bf2.7753e^{-1}$ &$\bf1.7652e^{-1}$ &$\bf7.9965e^{-2}$  &$\bf6.7721e^{-3}$  &$\bf1.5611e^{-4}$ \\

            \hline \hline
   \multirow{2}{*}{Damping}
            &$\beta$($L=1$) &$\bf{9.4710e^{-1}}$ &$ \bf9.3378e^{-1}$ &$\bf 9.6667e^{-1}$ &$\bf 7.1631e^{-1}$ &$\bf 3.7582e^{-1}$ \\
            &$\beta$($L=2$) &$\bf 8.8261e^{-1}$ &$\bf 9.3475e^{-1}$ &$\bf 9.1169e^{-1}$ &$\bf 4.8111e^{-1}$ &$\bf 2.5697e^{-1}$\\
            &$\beta$($L=3$) &$1.4997e^{-4}$ &$5.2261e^{-3}$ &$\bf 6.9141e^{-1}$ &$\bf 4.2961e^{-1}$ &$\bf 2.8293e^{-1}$\\
            &$\beta$($L=4$) &$1.2112e^{-4}$ &$1.2744e^{-5}$ &$\bf 1.3030e^{-1}$ &$\bf 3.4602e^{-1}$ &$\bf 1.3048e^{-1}$\\
            &$\beta$($L=5$) &$7.1442e^{-7}$ &$6.4718e^{-7}$ &$5.5887e^{-5}$  &$6.1929e^{-3}$ &$\bf 8.7179e^{-2}$ \\
            %\hline
               \bottomrule
  \end{tabular}
  \label{damping}
 \end{table}

Table \ref{damping} presents the performance comparisons between EP and EPNet for ${8 \times 8}$ uncoded MIMO systems with 16-QAM modulation. The performance of EPNet with 5 iterations is comparable with that of EP with 10 or 15 iterations. { The computation complexity of  EP and EPNet with 1 iteration is $\mathcal{O}(N_t^3)$ \cite{2014EP2}. } Given the same computational complexity (e.g., $L=5$), the BER performance of the trainable EP (i.e., EPNet) can be immensely improved. We list the learned damping factors in Table \ref{damping} to improve  the understanding of  the characteristics of EPNet. At a  low $E_B/N_0$ regime (e.g.,
$E_B/N_0$ $=-1$ and $4$ dB), the damping factors become extremely small after two iterations, implying that EPNet can reach convergence within two layers. Although EPNet requires many iterations to reach convergence at a high $E_B/N_0$ regime (e.g., $E_B/N_0$$=9$ and $14$ dB), it requires only four iterations. With the learned damping factors, we can determine the number of needed layers for EPNet, which is beneficial for reducing computational complexity. Notably, the damping factors vary with different $E_B/N_0$'s, indicating the importance of the online training mechanism because the environment constantly changes.

\begin{figure}[b]
\setlength{\abovecaptionskip}{-0.2 cm}
\setlength{\belowcaptionskip}{-0cm}
  \centering
  \includegraphics[width=3.5in]{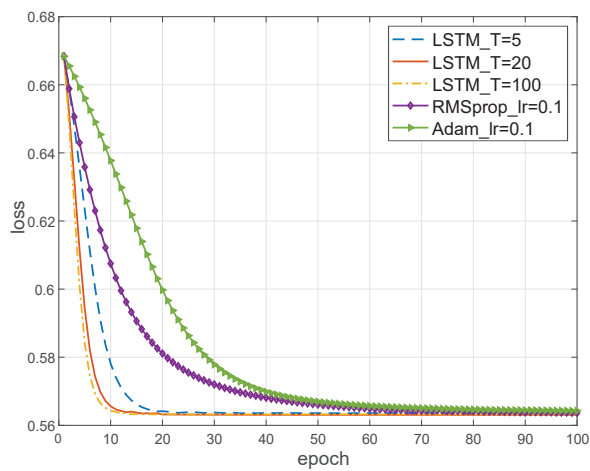}
        \caption{Convergence performance of the online training mechanism.}
  \label{online}
 \end{figure}

The convergence performance (loss versus epoch) of the online training mechanism is illustrated in Fig. \ref{online}. The loss value is the result of $f_{\rm EP}(\bm \beta)$. The convergence speed of EPNet depends on the time steps in the LSTM optimizer. As shown in Fig. \ref{online}, the convergence speed of EPNet with 5 time steps is evidently slower than that of EPNet with 20 time steps. However, no considerable  difference is observed with an  increase in time steps (i.e., $T$=100). This phenomenon suggests that the LSTM optimizer can find the optimal damping factors in a specific epoch with limited time steps. Therefore, the time steps of the LTSM optimizer are set as 20 in the remaining part of this study. EPNet can also be optimized using traditional optimizers, such as Adam and RMSprop, with a  learning rate of 0.1.  Its convergence speed performance is shown in Fig. \ref{online}. Adam and RMSprop optimizers spend 80 and 90 epochs to reach convergence. By contrast, the LSTM optimizer reaches convergence within 20 epochs. The convergence speed of the LSTM optimizer is evidently  faster than those of the  other optimizers, suggesting the high capability of meta learning-based EPNet in  adapting to new environments.

 \begin{table} [!b]
  \centering
  \footnotesize
        \caption{Complexity Comparison. }
  \begin{tabular}{lllllllll}    %lcrrr
   \toprule
   & Algorithm   & Complexity \\
                      %&(W)&  (W)  & (S) & (S) \\
   \midrule

&AMP/MMNet &$L\mathcal{O}(N_t^2)$ \\
&MMSE &$\mathcal{O}(N_t^3)$  \\
&OAMP/OAMPNet &$L\mathcal{O}(N_t^3)$  \\
&EP/EPNet &  $L\mathcal{O}(N_t^3)$\\
&ML &  $\mathcal{O}\left(|\mathcal{A}|^{N_{t}}\right)$\\

   \bottomrule
  \end{tabular}
  \label{complexity}
 \end{table}
\begin{figure}[!b]
\centering
\subfigure[Rayleigh channel in $32\times32$ MIMO systems.]{
\begin{minipage}[t]{0.46\textwidth}
\centering
\includegraphics[width=3in]{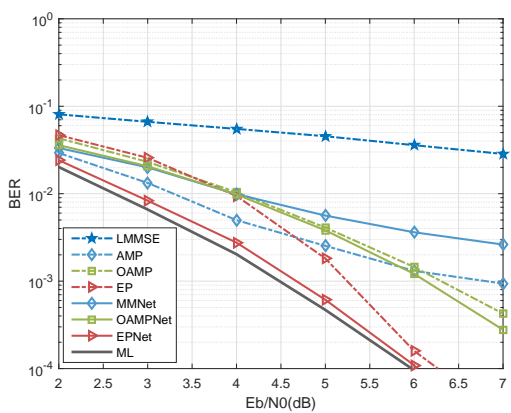}
%\caption{The impact of number of MTs.}
\label{fig4a}
\end{minipage}}
\subfigure[Correlated channel in $64\times32$ MIMO systems.]{
\begin{minipage}[t]{0.46\textwidth}
\centering
\includegraphics[width=3in]{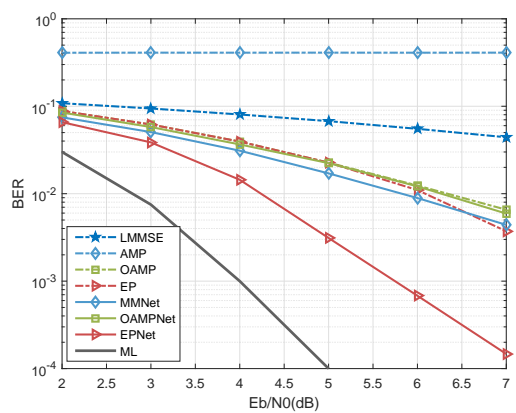}
%\caption{The impact of maximum tolerable delay.}
\label{fig4b}
\end{minipage}}
\caption{Performance comparison among existing detectors with QPSK modulation.}
\label{comparison}
\end{figure}

%%comparion
{  We compare  existing detectors with the proposed EPNet.   Jeon et al. \cite{2015amp} proposed an optimized AMP algorithm for MIMO detection, which  runs 50 iterations to reach convergence in our simulation. For the Rayleigh channel,  MMNet in the independent and identical distributed  case  \cite[(11)]{2019mmnet} is adopted. When the channel is correlated, MMNet for arbitrary channel matrices \cite[(13)]{2019mmnet} is used. The computation complexity of  AMP and MMNet is both $L\mathcal{O}(N_t^2)$.  OAMPNet in \cite{DBLP:journals/corr/abs-1809-09336}, which is based on  the OAMP algorithm \cite{2017OAMP}, is implemented in five layers with two trainable parameters per layer. The computation complexity of  the MMSE detector \cite{mmse2004} is dominated by the matrix inversion in (\ref{average}), which is $\mathcal{O}(N_t^3)$. The OAMP, OAMPNet, EP and EPNet with $L$ iterations exhibits an  $L$  matrix inversion operation, and thus, the computation complexity is $L\mathcal{O}(N_t^3)$.  ML with a  complexity of  $\mathcal{O}\left(|\mathcal{A}|^{N_{t}}\right)$ is an optimal baseline. The complexity comparison is concluded in Table \ref{complexity}. Fig. \ref{comparison} presents the performance comparison among existing detectors with QPSK modulation. As shown in the figure,  EPNet significantly outperform the existing algorithms either for the Rayleigh channel in $32\times32$ MIMO systems or the  3GPP 3D correlated channel \cite{2019mmnet} in $64\times32$ MIMO systems. In addition, EPNet with five iterations can nearly  achieve the performance of ML for the Rayleigh channel.

}

\begin{figure}[b]
\setlength{\abovecaptionskip}{-0.2cm}
\setlength{\belowcaptionskip}{-0.cm}
  \centering
  \includegraphics[width=3.5in]{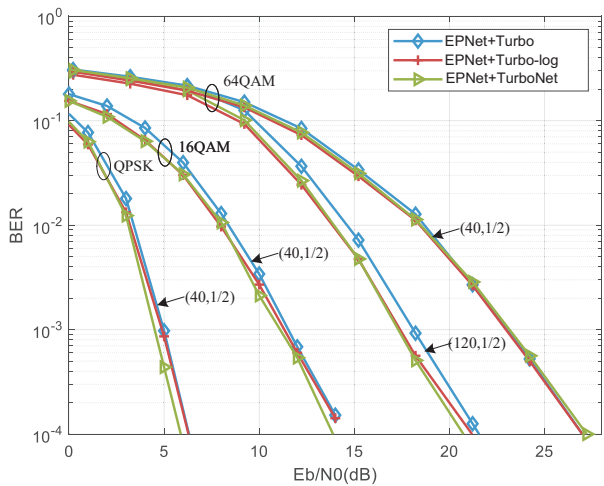}
        \caption{Performance comparison among different decoding algorithms  { at QPSK, 16-QAM and 64-QAM.}}
  \label{epnet+ep}
 \end{figure}

Fig. \ref{epnet+ep} shows the performance comparison among different turbo decoding algorithms, including max-log-MAP, log-MAP, and TurboNet, for QPSK, 16-QAM, and 64-QAM modulations. The computational complexity of TurboNet is comparable with  the max-log-MAP algorithm but lower than the log-MAP algorithm. In Fig. \ref{epnet+ep}, $(K,R)$ indicates that $K$ message bits are encoded by the turbo encoder at an $R$ rate. As shown in Fig. \ref{epnet+ep}, the performance of TurboNet is better than that of the max-log-MAP and comparable with  that of the log-MAP algorithm. The results indicate that TurboNet can achieve  better BER performance or lower computational complexity than the competing decoding algorithms. For $(40,1/2)$,  performance improvement can be easily observed for QPSK at 16-QAM. However,  improvement is inevident for 64-QAM because the code length is extremely short for  high-order modulation. The improvement of TurboNet for 64-QAM becomes more evident compared with that of the max-log-MAP algorithm when  code length increases (i.e., message bit length is 120). In particular, TurboNet is trained in AWGN channels with binary phase-shift keying modulation and low $E_b/N_0$ but is applied to Rayleigh fading channels with high-order modulation and high $E_b/N_0$, indicating the robustness of TurboNet. Consequently, TurboNet is trained \emph{off-line} only once because of this characteristic.
\subsection{Performance Analysis of the  Unfolded Turbo Receiver}
\begin{figure}[ht]
\setlength{\abovecaptionskip}{-0.2cm}
\setlength{\belowcaptionskip}{-0.cm}
  \centering
  \includegraphics[width=6in]{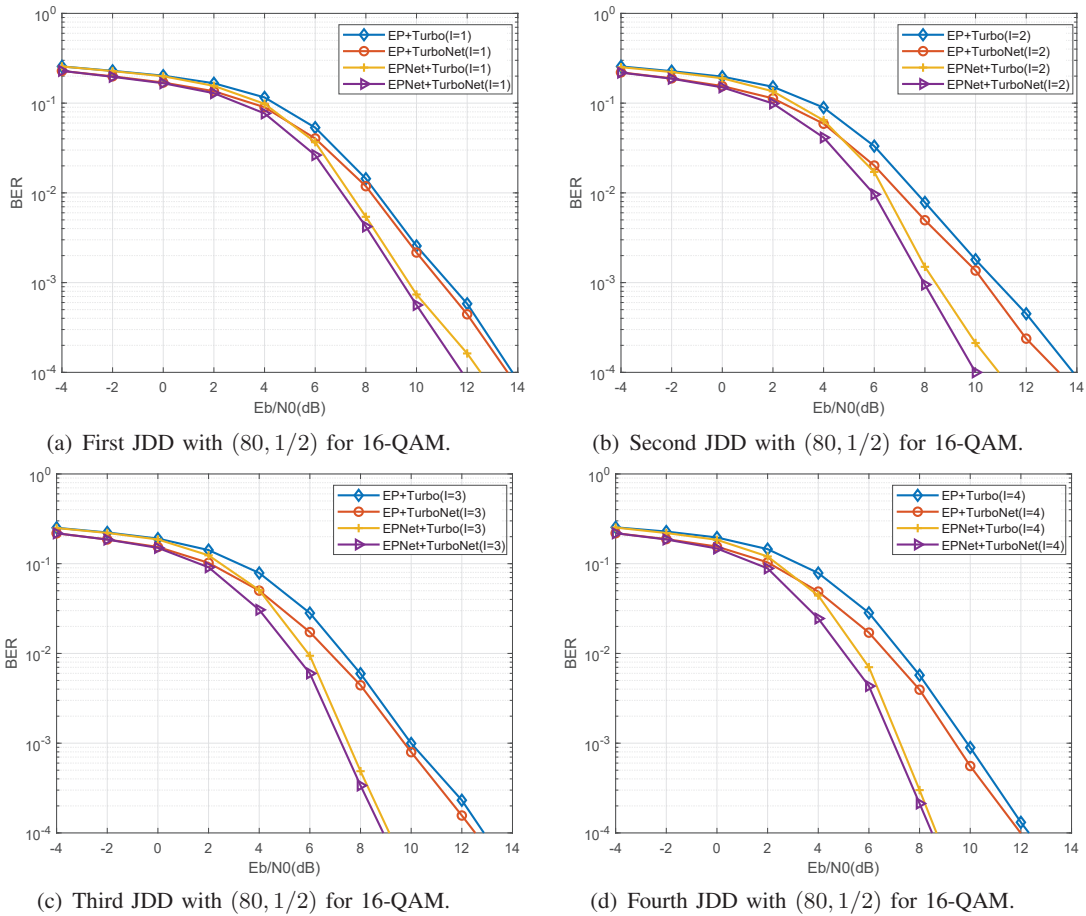}
        \caption{Performance comparison between different times of JDD with (80,1/2) for 16-QAM { in 32$\times$32 MIMO systems}.}
  \label{jdd}
 \end{figure}

\begin{table}[htbp]
  \centering
  \footnotesize
        \caption{ Damping factors trained by meta learning with 16-QAM in Rayleigh channel for 32$\times$32 MIMO systems}
  \begin{tabular}{>{\tt }lll|l|l|l|l|ll}    %lcrrr
   \toprule
    %&$E_b/N_0$  &  1.5 dB   &  2 dB       & 2.5 dB & 3 dB & 3.5 dB& 4 dB \\
&$E_b/N_0$  &  0 dB   &  2 dB       & 4 dB & 6 dB & 8 dB&10 dB \\

                      %&(W)&  (W)  & (S) & (S) \\
   \midrule
   \multirow{2}{*}{JDD-1}
            &$\beta$(L=1) &$\bf{9.4710e^{-1}}$ &$\bf{9.4710e^{-1}}$ &$ \bf9.3378e^{-1}$  &$\bf 9.5967e^{-1}$     &$\bf 9.7392e^{-1}$  &$\bf 8.2974e^{-1}$ \\
            &$\beta$(L=2)   &$ 8.8261e^{-3}$  &$ 8.8261e^{-3}$  &$ 9.3475e^{-3}$  &$\bf 9.4498e^{-1}$     &$\bf 9.5539e^{-1}$ &$\bf 8.4913e^{-1}$\\
            &$\beta$(L=3)   &$1.4997e^{-4}$      &$1.4997e^{-4}$      &$5.2261e^{-3}$      &$\bf 8.0388e^{-1}$     &$\bf 9.3365e^{-1}$  &$\bf 7.7950e^{-1}$\\
            &$\beta$(L=4)   &$1.2112e^{-4}$      &$1.2112e^{-4}$      &$1.2744e^{-5}$      &$ 6.3910e^{-3}$     &$ 3.8344e^{-3}$  &$\bf 7.9153e^{-1}$\\
            &$\beta$(L=5)   &$7.1442e^{-7}$      &$7.1442e^{-7}$   &$6.4718e^{-7}$    &$ 7.7625e^{-4}$    &$ 1.5600^{-4}$ &$\bf 7.3257e^{-1}$ \\
            \hline
            \multirow{2}{*}{JDD-2}
           &$\beta$(L=1) &$\bf8.5148e^{-1}$ &$\bf8.5148e^{-1}$ &$\bf8.6602e^{-1}$ &$\bf9.1322e^{-1}$ &$\bf8.1858e^{-1}$ &$\bf7.7651e^{-1}$\\
&$\beta$(L=2) &$2.3576e^{-3}$ &$2.3576e^{-3}$ &$2.5184e^{-3}$ &$\bf4.6149e^{-1}$ &$\bf5.5674e^{-1}$ &$\bf6.6807e^{-1}$\\
&$\beta$(L=3) &$1.5123e^{-3}$ &$1.5123e^{-3}$ &$1.6128e^{-3}$ &$\bf7.7709e^{-1}$ &$\bf3.1982e^{-1}$ &$\bf3.3266e^{-1}$\\
&$\beta$(L=4) &$1.0762e^{-3}$ &$1.0762e^{-3}$ &$1.1464e^{-3}$ &$\bf2.3704e^{-2}$ &$\bf1.7302e^{-1}$ &$\bf1.7729e^{-1}$\\
&$\beta$(L=5) &$8.04612e^{-4}$ &$8.04612e^{-4}$ &$8.5620e^{-4}$ &$\bf1.1824e^{-2}$ &$\bf1.0582e^{-1}$ &$\bf1.0798e^{-1}$\\
            \hline
            \multirow{2}{*}{JDD-3}
&$\beta$(L=1) &$\bf7.8859e^{-1}$ &$\bf8.5418e^{-1}$ &$\bf9.3443e^{-1}$ &$\bf9.0993e^{-1}$ &$\bf8.6756e^{-1}$ &$\bf7.8288e^{-1}$\\
&$\bf\beta$(L=2) &$3.6838e^{-3}$ &$2.7602e^{-3}$ &$\bf1.0837e^{-1}$ &$\bf9.3570e^{-1}$ &$\bf9.3569e^{-1}$ &$\bf7.2603e^{-1}$\\
&$\bf\beta$(L=3) &$2.4418e^{-3}$ &$1.8299e^{-3}$ &$\bf2.2887e^{-2}$ &$\bf8.8784e^{-1}$ &$\bf8.3987e^{-1}$ &$\bf5.6353e^{-1}$\\
&$\bf\beta$(L=4) &$1.8010e^{-3}$ &$1.3407e^{-3}$ &$6.2776e^{-3}$ &$\bf7.3046e^{-1}$ &$\bf8.0604e^{-1}$ &$\bf3.2636e^{-1}$\\
&$\bf\beta$(L=5) &$1.4596e^{-3}$ &$1.0180e^{-3}$ &$3.0399e^{-3}$ &$\bf2.9784e^{-1}$ &$\bf4.2794e^{-1}$ &$\bf1.4434e^{-1}$\\
            \hline
            \multirow{2}{*}{JDD-4}
&$\beta$(L=1) &$\bf9.0657e^{-1}$ &$\bf9.4442e^{-1}$ &$\bf9.8252e^{-1}$ &$\bf7.3057e^{-1}$ &$\bf9.8006e^{-1}$ &$\bf9.6234e^{-1}$\\
&$\bf\beta$(L=2) &$\bf1.1644e^{-1}$ &$\bf1.1880e^{-1}$ &$\bf3.6986e^{-1}$ &$\bf5.7898e^{-1}$ &$\bf8.8634e^{-1}$ &$\bf7.5010e^{-1}$\\
&$\bf\beta$(L=3) &$\bf5.1860e^{-2}$ &$\bf3.8813e^{-2}$ &$\bf8.9909e^{-2}$ &$\bf4.1034e^{-1}$ &$\bf9.4007e^{-1}$ &$\bf3.0014e^{-1}$\\
&$\bf\beta$(L=4) &$\bf2.1213e^{-2}$ &$\bf1.1810e^{-2}$ &$\bf3.3124e^{-2}$ &$\bf2.2581e^{-1}$ &$\bf8.8783e^{-1}$ &$\bf1.6493e^{-1}$\\
&$\bf\beta$(L=5) &$8.2202e^{-3}$ &$4.4983e^{-3}$ &$\bf1.5950e^{-2}$ &$\bf1.1998e^{-1}$ &$\bf2.5634e^{-1}$ &$\bf1.0649e^{-1}$\\
               \bottomrule
  \end{tabular}
  \label{dampings4}
 \end{table}

Fig. \ref{jdd} shows the BER versus $E_b/N_0$ performance of the unfolded turbo receiver with ${32 \times 32}$ antennas, 16-QAM modulation, and turbo code rate of 1/2. BER performance improves with an increase in the number of iterations between the detector and the decoder. The performance gaps among the first, second, and third JDDs are considerably apparent. However, the performance gap between the third and fourth JDDs decreases, indicating that performing more than four iterations does not compensate considerably  in practical systems.

The learned values are listed in Table \ref{dampings4} to investigate the damping factors of EPNet for each JDD. The optimal damping factors of EPNet for each layer are different. A layer with an extremely small damping factor is unnecessary. Therefore, the required number of  layers of EPNet increases with an  increase in $E_b/N_0$, as shown in Table \ref{dampings4}. This condition is attributed to  the interior of EPNet passing  only  unreliable information and failing  to  improve the EP performance by increasing the layers at low $E_b/N_0$. However, it is effective at high $E_b/N_0$. In addition, the required number of  layers of EPNet increases with an increase in iterations under the same $E_b/N_0$. This condition is reasonable because  BER performance significantly improves with the increase in JDD iterations and is comparable with  that at high $E_b/N_0$. Moreover, the variation range of the damping values with $E_b/N_0 < 6$ dB is smaller than that with $E_b/N_0 \ge 6$ dB for different JDD iterations. For high $E_b/N_0$, the damping values of {\tt JDD}$-1$, {\tt JDD}$-2$, and {\tt JDD}$-3$ significantly vary. By contrast,   the difference between {\tt JDD}$-3$ and {\tt JDD}$-4$ decreases, which is consistent with the slight  performance improvement  shown in Fig. \ref{jdd}(d) compared with that in Fig. \ref{jdd}(c).

\begin{figure}[ht]
\setlength{\abovecaptionskip}{-0.8cm}
\setlength{\belowcaptionskip}{-0.cm}
  \centering
  \includegraphics[width=6.8in]{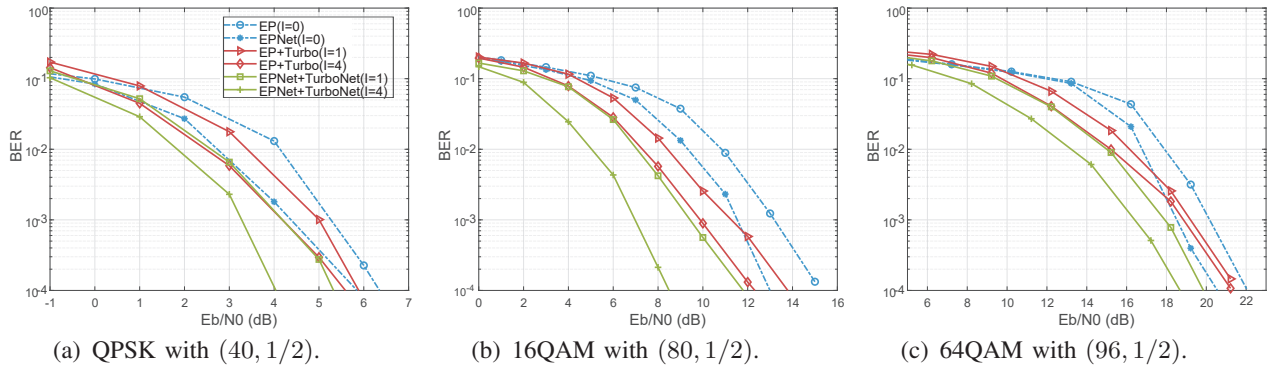}
        %\caption{Performance comparison among QPSK, at 16-QAM and 64-QAM for ${32 \times 32}$ MIMO system.}
        \caption{Performance comparison { at QPSK,  16-QAM and 64-QAM} for ${32 \times 32}$ MIMO systems.}
 \label{4_16_64-QAM}
 \end{figure}

Fig. \ref{4_16_64-QAM} shows the performance comparison of different receivers for QPSK with $(40, 1/2)$, 16-QAM with $(80, 1/2)$, and 64-QAM with $(96, 1/2)$ in a ${32 \times 32}$ MIMO system. To ensure fairness, the $E_B/N_0$ of the uncoded receivers is 3 dB more than the $E_b/N_0$ of the coded receivers in Fig. \ref{4_16_64-QAM}. EPNet can significantly improve  BER performance compared with EP for all the simulated modulations. As shown in Fig.  \ref{4_16_64-QAM}(a), the performance of EPNet is comparable with that of EP+Turbo ($I=4$), demonstrating the importance of damping factor settings. For the turbo receiver, EPNet+TurboNet, with only one iteration (i.e., $I=1$), is comparable with EP+Turbo ($I=4$). The performance of EPNet+TurboNet can be immensely improved with an  increase in the number of iterations. In summary, compared with EP+Turbo ($I=4$), EPNet+TurboNet ($I=4$) provides additional 2-, 4-, and 3-dB gains for QPSK, 16-QAM, and 64-QAM, respectively.

\section{OTA Test and Result Discussion}
In this section, a prototyping system is presented to verify the effectiveness and feasibility of the proposed unfolded turbo receiver in real channel environments.

\subsection{System Setup}

\begin{figure}[t]

  \centering
  \includegraphics[width=4.5in]{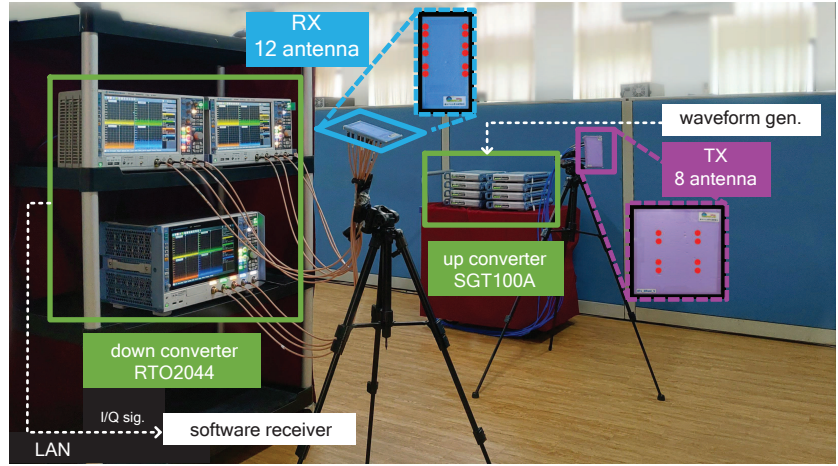}
        \caption{Diagram of system configuration.}
  \label{system_configuration}
 \end{figure}

1) Prototyping platform configuration: As shown in Fig. \ref{system_configuration}, the platform is a ${12 \times 8}$ MIMO OFDM system operating at a 3.5 GHz frequency band. The base station, as the transmitter, uses four dual-polarized patch antennas that  can support eight radio frequency (RF) channels of simultaneous data transmission( known as 8-port patch antennas). A mobile phone, as the receiver, uses 12 compact antennas that   can simultaneously receive 12 RF channels. The antennas are placed at the upper half of the two long sides (146 mm) of the mobile phone. In the case without  isolation components, six antennas are placed at approximately $0.8$ wavelength, and the antenna envelope correlation coefficient is less than $0.1$. The platform can establish a ${12 \times 8}$ or ${8 \times 8}$ MIMO system. The detailed transmitter and receiver configurations are described as follows.

\begin{itemize}
\item Transmitter configuration: The signal generator (SGT100A) of Rohde and  Schwarz (R\&S) is used. The transmitter first generates a transmission waveform and then uses SGT100A to modulate the signal from the fundamental frequency to the frequency band of 3.5 GHz (the hot band of 5G). The RF signal is emitted by one of the transmit antennas.

\item Receiver configuration: After the signal is received by the receive antennas, the R\&S digital oscilloscope (RTO2044) is used to convert the analog signal into a digital signal and demodulate it to the fundamental frequency. Then, the oscilloscope sends the in-phase/quadrature signal to the computer via a local area network  to perform demodulation (called the software receiver). The software receiver performs signal synchronization, channel estimation, signal detection, and channel decoding operations to recover the data. The unfolded turbo receiver and the online training mechanism are deployed in the software receiver. {  In our OTA test, the software receiver is implemented on one computer with an Intel CPU (1.60 GHz frequency, 16 GB  memory and 180 W computation power).  The authors of  \cite{2018ISSCCep} presented a 2.0mm$^2$  $128\times16$ massive MIMO detector application specific integrated circuit that  implements iterative EP detection  and  enables 2.7$\times$ reduction in  power transmmition for battery-powered mobile terminals.  As analyzed in the Section IV. B, our meta learning-based EPNet has lower computation complexity than EP and can reach convergence within 20 training epochs. In addition,  EPNet is highly relevant with  channel statistics and not required  to be retrained unless the channel environment considerably  changes.  Therefore,  implementing  the proposed framework in  mobile devices with limited power and computation capability is feasible.} In the experiment environment, TurboNet and the LSTM optimizer are trained off-line the same as that in the simulation.
\end{itemize}

\begin{figure}[ht]
\setlength{\abovecaptionskip}{0.cm}
\setlength{\belowcaptionskip}{0cm}
  \centering
  \includegraphics[width=6in]{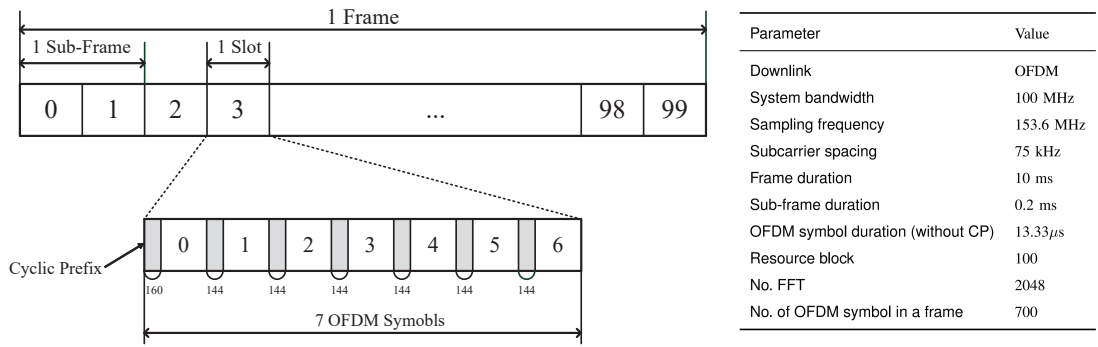}
        \caption{LTE frame structure.}
  \label{frame_structure}
 \end{figure}

2) Frame structure: OFDM is used in the prototyping system,  and the frame structure of the transmission data specification is modified from the Long-Term Evolution (LTE) standard. The modified frame structure is in the bandwidth of 100 MHz, and a complete frame time length is 10 ms. As shown in Fig. \ref{frame_structure}, each frame contains 50 equal-length sub-frames that comprise two time slots. Each time slot comprises seven OFDM symbols and uses a normal CP of LTE. The CP length of the first symbol is 160 sampling points and that of the other symbols is 144 sampling points. The relevant parameters are also shown in Fig. \ref{frame_structure}. Pilot symbols are inserted into time and frequency similar to those used in LTE to facilitate channel estimation.

%\vspace{-1em}
\subsection{Indoor and Outdoor Scenarios}

\begin{figure}[t!]

  \centering
  \includegraphics[width=4.2in]{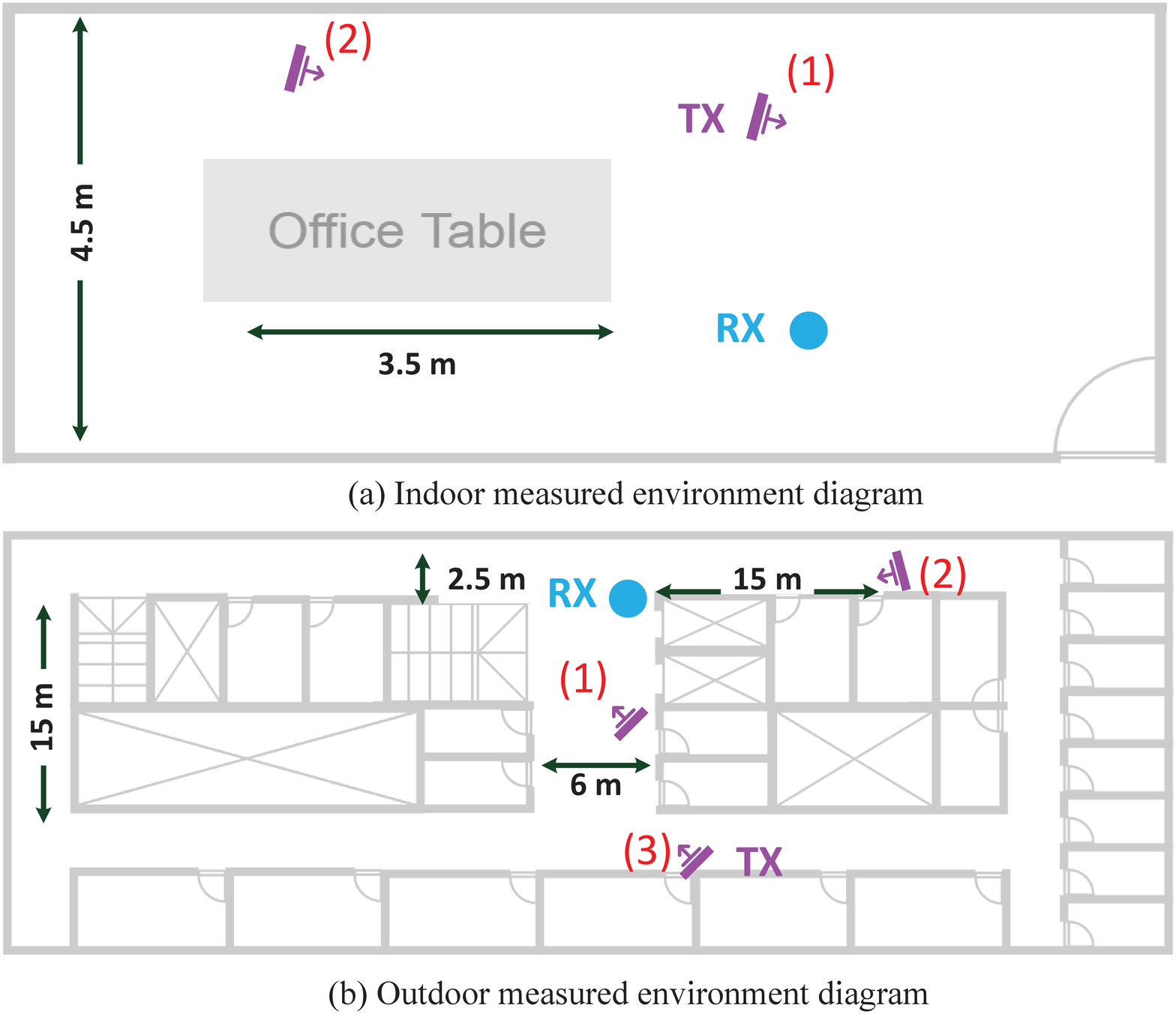}
        \caption{{ Diagrams of indoor  and outdoor} measured environments.}
  \label{indoor_environment}
 \end{figure}

In the OTA test, the platform is used to collect  test data in different times and environments. Data collection is  divided into indoor and outdoor. Indoor refers to a laboratory, and outdoor refers to a corridor. Different scenarios in each environment (indoor and outdoor) are considered.
One scenario is selected as the benchmark, and the remaining of the scenarios are compared with the adjustment on the benchmark.
{ The major indoor and outdoor measurement scenarios are described in Table \ref{indoor-s} (Fig. \ref{indoor_environment}).  The corresponding indoor measurement scenarios are shown in Fig. \ref{indoor_environment}(a), and the major outdoor measurement scenarios are presented as Fig. \ref{indoor_environment}(b).}

 \begin{table} [!h]
  \centering
  \footnotesize
        \caption{Indoor and Outdoor Measurement Scenarios.}
  \begin{tabular}{>{\sf }ll|ll|l|llll}    %lcrrr
   \toprule
   & Scenarios   & Description \\
                      %&(W)&  (W)  & (S) & (S) \\
   \midrule

   \multirow{6}{*}{Indoor}
&s-i &  Place the transmitter in position (1) and collect data as the benchmark.\\
&s-ii &  Place in the same position as the benchmark, and a person is walking. \\
&s-iii &  Place the data in the same position as the benchmark and measure the data after 30 min. \\
&s-iv&  Change the angle of the transmit antenna at the same position as the benchmark.\\
&s-v & Change another angle of the transmitting antenna different from iv. \\
&s-vi&  Place the transmitter in position (2) and measure the long distance scenario.\\
                      %&(W)&  (W)  & (S) & (S) \\
      \hline
   \hline
            \multirow{6}{*}{outdoor}
&s-i &  Place the transmitter in position (1) in a wide corridor environment as the benchmark.\\
&s-ii &  Place in the same position as the benchmark, and a person is walking.\\
&s-iii &  Place the data in the same position as the benchmark and measure the data after 30 min. \\
&s-iv&  Change the angle of the transmit antenna at the same position as the benchmark.\\
&s-v & Place the transmitter in position (2) in the narrow corridor.\\
&s-vi&  Place the transmitter in position (3)  without direct vision between the transceiver antennas.\\

   \bottomrule
  \end{tabular}
  \label{indoor-s}
 \end{table}

In the experiment, the channel decoding schemes are the same as the settings shown in Fig. \ref{4_16_64-QAM}. In particular, $(40,1/2)$, $(80,1/2)$, and $(96,1/2)$ are applied to QPSK at 16-QAM and 64-QAM. Under the same position, we vary the transmit power to obtain different received SNRs for reflecting  performance under various  coding and modulation schemes. In a practical set-up, the received SNRs for each antenna are different. In the following experiments, all the turbo receivers perform two JDD iterations, i.e., $I=2$. The performance of EPNet+TurboNet(s-i) represents the performance of the proposed turbo receiver that is trained under the baseline scenario (s-i) and tested on  other scenarios (s-ii to s-vi). Meanwhile, EPNet+TurboNet denotes that the proposed turbo receiver is trained and tested in the corresponding scenario.

%%%%%%%%%%%1111111111111
%indoor ${8 \times 8}$
\begin{table} %[!b]
  \centering
  \footnotesize
        \caption{BER Comparison for ${8 \times 8}$ MIMO System in the Indoor Environment.}
  \begin{tabular}{>{\sf }lll|l|l|l|l|llll}    %lcrrr
   \toprule
   Modulation    &Algorithm  &  i: TX (1)    & ii: walking  & iii: 30 min  &  iv: angle 1    & v: angle 2  & vi: TX (2) \\
                      %&(W)&  (W)  & (S) & (S) \\
   \midrule
   \multirow{5}{*}{QPSK}
&EP &  $2.3250e^{-2}$ &  $1.0323e^{-2}$ &  $2.0479e^{-2}$ &  $5.3078e^{-2}$ &  $2.1505e^{-2}$ &  $7.9354e^{-2}$\\
&EPNet &  $1.4490e^{-2}$ &  $4.3802e^{-3}$ &  $1.1510e^{-2}$ &  $3.3958e^{-2}$ &  $1.1849e^{-2}$ &  $6.0354e^{-2}$\\
&EP+Turbo &  $3.1550e^{-3}$ &  $1.3000e^{-3}$ &  $1.6550e^{-3}$ &  $1.4860e^{-2}$ &  $4.3450e^{-3}$ &  $3.6640e^{-2}$\\
&EP+TurboNet &  $2.6350e^{-3}$ &  $1.0550e^{-3}$ &  $9.5500e^{-4}$ &  $1.1290e^{-2}$ &  $3.7100e^{-3}$ &  $2.8835e^{-2}$\\
&{EPNet+TurboNet(s-i)} &  $\bf1.4450e^{-3}$ &  $\bf3.8500e^{-4}$ &  $\bf7.3500e^{-4}$ &  $\bf7.1900e^{-3}$ &  $\bf1.3100e^{-3}$ &  $\bf1.7630e^{-2}$\\
&{EPNet+TurboNet} &  $\bf1.4450e^{-3}$ &  $\bf7.5000e^{-5}$ &  $\bf6.8500e^{-4}$ &  $\bf5.7950e^{-3}$ &  $\bf1.0250e^{-3}$ &  $\bf1.6420e^{-2}$\\

            \hline
            \multirow{2}{*}{16-QAM}
&EP &  $3.8705e^{-2}$ &  $3.3186e^{-2}$ &  $3.3531e^{-2}$ &  $7.1294e^{-2}$ &  $4.9815e^{-2}$ &  $9.2120e^{-2}$\\
&EPNet &  $2.4970e^{-2}$ &  $2.0100e^{-2}$ &  $2.1090e^{-2}$ &  $5.2485e^{-2}$ &  $3.2707e^{-2}$ &  $8.0552e^{-2}$\\
&EP+Turbo &  $9.0649e^{-4}$ &  $1.1147e^{-3}$ &  $3.7734e^{-4}$ &  $1.1255e^{-2}$ &  $4.6842e^{-4}$ &  $9.5767e^{-3}$\\

&{EPNet+TurboNet(s-i)} &  $\bf3.8602e^{-4}$ &  $\bf8.6312e^{-4}$ &  $\bf2.6024e^{-4}$ &  $\bf8.9001e^{-3}$ &  $\bf3.9903e^{-4}$ &  $\bf6.4885e^{-3}$\\
&{EPNet+TurboNet} &  $\bf3.8602e^{-4}$ &  $\bf3.0722e^{-4}$ &  $\bf2.2554e^{-4}$ &  $\bf6.0939e^{-3}$ &  $\bf3.0795e^{-4}$ &  $\bf4.3156e^{-3}$\\
            \hline
            \multirow{3}{*}{64-QAM}
&EP &  $3.8705e^{-2}$ &  $3.3186e^{-2}$ &  $3.3531e^{-2}$ &  $7.1294e^{-2}$ &  $4.9815e^{-2}$ &  $9.2120e^{-2}$\\
&EPNet &  $9.3587e^{-3}$ &  $8.1707e^{-3}$ &  $1.0898e^{-2}$ &  $1.1253e^{-2}$ &  $7.4857e^{-3}$ &  $3.6479e^{-2}$\\
&EP+Turbo &  $1.3939e^{-2}$ &  $1.2422e^{-2}$ &  $1.3555e^{-2}$ &  $1.4102e^{-2}$ &  $9.3544e^{-3}$ &  $2.1749e^{-2}$\\
&{EPNet+TurboNet(s-i)} &  $\bf2.5852e^{-3}$ &  $\bf1.5121e^{-3}$ &  $\bf2.5418e^{-3}$ &  $\bf1.6205e^{-3}$ &  $\bf6.8288e^{-4}$ &  $\bf1.0986e^{-2}$\\
&{EPNet+TurboNet} &  $\bf2.5852e^{-3}$ &  $\bf8.6715e^{-4}$ &  $\bf1.9457e^{-3}$ &  $\bf1.1815e^{-3}$ &  $\bf5.7019e^{-4}$ &  $\bf8.2976e^{-3}$\\
   \bottomrule
  \end{tabular}
  \label{indoor88}
 \end{table}

Table \ref{indoor88} provides the corresponding BER results in indoor scenarios. Given a receiver, the BER performance between scenarios s-i and s-iii are close. Meanwhile, the BER performance among different scenarios are relatively different when the transceiver is located in the same place but with different angle rotations at the transmitter/receiver side. These results indicate that  indoor conditions are approximately time-invariant but location-variant. Among the different receivers, EPNet+TurboNet is the best, followed by EPNet+TurboNet(s-i) and EP+TurboNet. EPNet+TurboNet(s-i) performs better than EP+TurboNet because of the introduction of learnable damping factors. Meanwhile, EPNet+TurboNet performs better than EPNet+TurboNet(s-i) because of its additional  computational costs to train  damping factors for each scenario.  In some cases (e.g., scenarios s-iii and s-v), the improvements between EPNet+TurboNet and {EPNet+TurboNet(s-i)} are insignificant, demonstrating the robustness of the proposed turbo receiver to a certain  extent. The experiment results suggest that the proposed turbo receiver exhibits considerable  robustness to the environment and does not require  frequent training. Online training is necessary  when the best performance is required  in a specific scenario. For 64-QAM, the BER performance\footnote{Note that we use normalized $E_B/N_0$ to make a fair comparison between uncoding and coded systems.} of EPNet is slightly better than that of EP+Turbo in scenarios s-i to s-v, indicating the benefit of training  damping factors. EP+TurboNet is significantly better than  EP+Turbo. The results demonstrate the usefulness and robustness of TurboNet in a practical MIMO channel.
\begin{table} %[t!]
  \centering
  \footnotesize
        \caption{BER Comparison for ${12 \times 8}$ MIMO System in the Outdoor Environment.}
  \begin{tabular}{>{\sf }lll|l|l|l|l|llll}    %lcrrr
   \toprule
   Modulation    &Algorithm  &  { i}: TX (1)    &{ ii}: walking  & {iii}: 30 min  &  { iv}: angle 1    &{ v}: TX (2)  & {vi}: TX (3) \\
   \midrule
   \multirow{2}{*}{QPSK}
&EP &  $2.8160e^{-2}$ &  $2.5020e^{-2}$ &  $2.4922e^{-3}$ &  $6.9961e^{-3}$ &  $3.1574e^{-2}$ &  $1.5363e^{-1}$\\
&EPNet &  $1.5297e^{-2}$ &  $1.3215e^{-2}$ &  $8.3594e^{-4}$ &  $3.3711e^{-3}$ &  $1.9633e^{-2}$ &  $1.4720e^{-1}$\\
&EP+Turbo &  $2.2036e^{-3}$ &  $2.8250e^{-3}$ &  $1.0607e^{-3}$ &  $1.7071e^{-3}$ &  $6.1571e^{-3}$ &  $1.8285e^{-1}$\\
&{EPNet+TurboNet(s-i)} &  $\bf1.7464e^{-3}$ &  $\bf2.2964e^{-3}$ &  $\bf2.8214e^{-4}$ &  $\bf8.4286e^{-4}$ &  $\bf4.1679e^{-3}$ &  $\bf1.6813e^{-1}$\\
&{EPNet+TurboNet} &  $\bf1.7464e^{-3}$ &  $\bf1.9000e^{-3}$ &  $\bf2.1429e^{-4}$ &  $\bf7.9643e^{-4}$ &  $\bf3.2321e^{-3}$ &  $\bf1.3437e^{-1}$\\
            \hline
            \multirow{2}{*}{16-QAM}
&EP &  $1.4029e^{-1}$ &  $1.3737e^{-1}$ &  $6.6841e^{-2}$ &  $7.4366e^{-2}$ &  $1.4254e^{-1}$ &  $2.4693e^{-1}$\\
&EPNet &  $1.3387e^{-1}$ &  $1.2992e^{-1}$ &  $5.5897e^{-2}$ &  $6.5169e^{-2}$ &  $1.3560e^{-1}$ &  $2.4273e^{-1}$\\
&EP+Turbo &  $9.3950e^{-2}$ &  $8.6962e^{-2}$ &  $1.0015e^{-2}$ &  $1.7002e^{-2}$ &  $1.1473e^{-1}$ &  $2.2746e^{-1}$\\
&{EPNet+TurboNet(s-i)} &  $\bf6.1914e^{-2}$ &  $\bf6.9175e^{-2}$ &  $\bf5.9160e^{-3}$ &  $\bf1.0934e^{-2}$ &  $\bf9.4696e^{-2}$ &  $\bf2.2253e^{-1}$\\
&{EPNet+TurboNet} &  $\bf6.1914e^{-2}$ &  $\bf5.3587e^{-2}$ &  $\bf4.0727e^{-3}$ &  $\bf6.9613e^{-3}$ &  $\bf7.4675e^{-2}$ &  $\bf1.9317e^{-1}$\\

            \hline
            \multirow{3}{*}{64-QAM}
&EP &  $1.2203e^{-2}$ &  $8.8580e^{-3}$ &  $4.9644e^{-4}$ &  $1.3701e^{-3}$ &  $8.3117e^{-3}$ &  $1.3541e^{-1}$\\
&EPNet &  $6.4364e^{-3}$ &  $4.6848e^{-3}$ &  $1.6259e^{-4}$ &  $6.5904e^{-4}$ &  $4.5937e^{-3}$ &  $1.3052e^{-1}$\\
&EP+Turbo &  $4.4604e^{-3}$ &  $3.3873e^{-3}$ &  $5.8533e^{-4}$ &  $5.9617e^{-4}$ &  $2.0270e^{-3}$ &  $1.9972e^{-1}$\\
&{EPNet+TurboNet(s-i)} &  $\bf1.3820e^{-3}$ &  $\bf6.9914e^{-4}$ &  $\bf1.4091e^{-5}$ &  $\bf1.7343e^{-4}$ &  $\bf4.7151e^{-4}$ &  $\bf1.7631e^{-1}$\\
&{EPNet+TurboNet} &  $\bf1.3820e^{-3}$ &  $\bf4.7151e^{-4}$ &  $\bf5.4783e^{-6}$ &  $\bf1.2465e^{-4}$ &  $\bf1.5175e^{-4}$ &  $\bf1.4396e^{-1}$\\
   \bottomrule
  \end{tabular}
  \label{outdoor128}
 \end{table}

A ${12 \times 8}$ MIMO system is tested in an  outdoor environment to verify the robustness of the proposed scheme. Table \ref{outdoor128} provides  the corresponding results. Similar insights can be derived from  the indoor ${8 \times 8}$ MIMO system. The proposed turbo receiver demonstrates   considerable  robustness and outperforms traditional turbo receivers. Overall performance sharply deteriorates  in scenario s-vi. In this scenario, the transmitter is in the corner of the wall, and the multipath is too  weak to support eight spatial multiplexing streams.

\section{Conclusion}

We propose an efficient online training framework by separating an NN-based MIMO receiver into channel-sensitive and channel-insensitive modules. To achieve  a systematic design, we use a deep unfolded NN structure that represents  iterative algorithms in signal detection and channel decoding modules as multi-layer NNs. In particular, the EP algorithm for signal detection is unfolded as EPNet, and the damping factors are set as trainable parameters to adapt to new channels. An unfolded turbo decoding module, TurboNet, is used in  channel decoding. TurboNet is robust for channels. Therefore, it  only requires to be trained once in  an off-line setting, and only EPNet needs to be trained online. Damping factors are relevant to  channel statistics rather than instantaneous channel realizations, and thus, the training labels can be easily obtained by generating similar channels locally at the receiver side. We further develop an online training mechanism based on meta learning; here, the LSTM optimizer is trained to  update the damping factors efficiently by using a small training set such  that they can quickly adapt to new environments.
The simulation and OTA test results indicate  that the deep unfolded turbo receiver immensely outperforms traditional turbo receivers and is extremely  robust. Moreover,  the online training mechanism demonstrates  high  capability to adapt to new  environments.
\section*{Acknowledgement}
The authors would like to thank the NSYSU Antenna Laboratory led by Prof. K. L. Wong to design the antennas at the mobile phone and base station for the OTA testing.

% \bibliographystyle{IEEEtran}
%    %\bibliography{IEEEabrv,metabib0909}
%    \bibliography{IEEEabrv,new0306}

\end{document}